\documentclass[12pt,a4paper]{article}
\usepackage[utf8]{inputenc}
\usepackage[english]{babel}
\usepackage[T1]{fontenc}
\usepackage{amsmath}
\usepackage{amsfonts}
\usepackage{amssymb}
\usepackage{graphicx}
\usepackage{lmodern}

\usepackage[left=2cm,right=2cm,top=2cm,bottom=2cm]{geometry}
\usepackage{fancyhdr}
\usepackage{enumerate}

\usepackage{amssymb}
\usepackage{amsfonts}
\usepackage{amsmath}
\usepackage{perpage}
\MakePerPage{footnote}
\usepackage[titletoc,title]{appendix}

\begin{document}
\pagestyle{fancy}
\title{Covariant entropic dynamics: from path independence to Hamiltonians and quantum theory}
\author{Selman Ipek\\
University at Albany\footnote{sipek@albany.edu}}
\date{}
\maketitle

\begin{abstract}
Entropic Dynamics (ED) is an inference-based framework that seeks to  construct dynamical theories of physics without assuming the conventional formalism --- the Hamiltonians, Poisson brackets, Hilbert spaces, etc. --- typically associated with physics. In this work we develop an ED of scalar fields that is both quantum and manifestly covariant. The framework for accomplishing this is inspired by the covariant methods of Dirac, Teitelboim, and Kucha\v{r}. In addition to the ostensible result of a covariant quantum ED, we also show how the covariance requirement of \textit{path independence} proposed by Teitelboim is sufficient for a \textit{derivation} of Hamiltonian dynamics and also provides a proof of uniqueness for the quantum potential that leads to quantum theory.
\end{abstract}

\section{Introduction}
In entropic dynamics (ED) quantum theory is derived as a theory of dynamical probability $\rho=|\Psi|^{2}$ and drift potential $\phi$ by the application of inference-based methods, i.e. probability theory and Maximum Entropy (ME). This viewpoint establishes a clear interpretation for the quantum state $\Psi$ as relating to a state of knowledge --- an \textit{epistemic} state. The challenge then in ED is to establish an updating scheme (i.e. a dynamics) for the variables $\rho$ and $\phi$ that is consistent with $\rho=|\Psi|^{2}$ being a Bayesian probability, but one that also, of course, agrees with the conventional quantum dynamics: the Schr\"{o}dinger equation. To this point, while the update of the probability $\rho$ must be consistent with the established Bayesian and entropic methods for updating,\footnote{These methods are based on the pioneering work of R.T. Cox \cite{Cox} and E.T. Jaynes \cite{Jaynes}, which were later refined and unified by A. Caticha and A. Giffin \cite{Caticha/Giffin 2006}; for more details on this unification see the thesis of A. Giffin \cite{Giffin}; and for a historical overview of Bayesian and entropic methods, as well as a demonstration of their unification see \cite{Caticha 2012}.} the criterion for updating the drift potential $\phi$, on the other hand is not prescribed purely by the rules of inference. The evolution of $\phi$ is instead driven by the constraints that reflect the relevant physics. It is this non-trivial step, however --- the identification of the appropriate constraints --- that requires special care.

In previous iterations of ED \cite{Caticha Ipek 2014}\cite{Bartolomeo et al 2014}, formulated on a flat space, one could utilize a scheme, first devised within the context of Nelson's stochastic mechanics \cite{Nelson}, for updating the drift potential $\phi$. There, the appropriate information for updating $\phi$ was so that a \emph{global} energy functional would be conserved. The result was a non-dissipative ED where $\rho$ and $\phi$ were cast as a canonical pair\footnote{While strictly speaking $\rho$ and $\phi$ do form a canonical pair, the theory is best formulated using the variable $\Phi$, which is to be introduced later.} within a Hamiltonian framework, together with the canonical structure that this normally entails, i.e. symplectic structure, Poisson brackets, Hamiltonian generators, etc. To go further towards recovering quantum dynamics, however, required, in addition, a special choice of potential --- the so-called \textit{quantum potential}. Contingent on this choice of potential, the unitary Schr\"{o}dinger time evolution and patented linearity of quantum theory followed.

This prescription is quite legitimate if one is dealing with a flat space (as was the case in \cite{Caticha Ipek 2014}\cite{Bartolomeo et al 2014}), since the translational symmetry makes it natural to introduce a \emph{global} time, and thus a global Hamiltonian generator is valid. In curved spacetime, however, the notion of a globally conserved energy fails completely and a new updating protocol must be devised. One would expect that such an alternative rule for updating $\phi$ would discard the global criterion of a conserved energy in favor of a \emph{local} rule for updating, which could be implemented in curved spacetime as well.

A similar problem occurs with regards to the role of time in ED, which is particularly apparent in \cite{Caticha 2013}\cite{Caticha Ipek 2014}. There, ED was used to success in developing a quantum theory of relativistic scalar fields in the Schr\"{o}dinger representation. Indeed, although the result there was a legitimate relativistic quantum field theory, the theory failed in depicting the arbitrary notions of simultaneity that are quintessential to relativity. That is, even though the theory was fully relativistic, the invariance of the theory was not explicit, or \emph{manifest}.\footnote{In fact, the theory in \cite{Caticha Ipek 2014} failed in much the same way that early attempts from the founders of quantum theory failed when they similarly tried to quantize relativistic field theories: the approaches were not manifestly covariant. This had largely to do with the fact that these attempts had borrowed wholesale the quantization schemes that Dirac, Heisenberg, Schr\"{o}dinger, etc., had developed for non-relativistic particles. The situation was remedied somewhat by the early many-times approaches of Weiss \cite{Weiss 1938} and Dirac \cite{Dirac 1932}. These approaches were later made fully covariant by Tomonaga \cite{Tomonaga 1946} and Schwinger \cite{Schwinger 1948 I}, who utilized them in their Nobel prize works. In turn, the formalism resulting from these early attempts eventually became the precursor to Dirac's \cite{Dirac 1951}\cite{Dirac Lectures} later works on generalized Hamiltonian dynamics, which then paved the way for the work of Teitelboim \cite{Teitelboim 1972}\cite{Teitelboim thesis} and Kucha\v{r} \cite{Kuchar 1972}.}

And so, advances in the ED approach to quantum field theory have been hindered by two main roadblocks:
\begin{itemize}
\item[1.] The ability to construct a locally covariant updating procedure for $\rho$ and $\phi$ that is valid in generic spacetimes.
\item[2.] A rigorous argument for the quantum potential that does not rely strictly on heuristic reasoning.
\end{itemize}
While one would expect \emph{a priori} that these problems are completely unrelated, the surprising conclusion of this work is that, in fact, they are not: requiring ED to be manifestly covariant comes with a set of constraints that, when satisfied, single out the quantum potential, nearly uniquely.

The methods for accomplishing this are inspired by the works of Dirac \cite{Dirac 1951}\cite{Dirac Lectures}, Kucha\v{r} \cite{Kuchar 1972}, and Teitelboim \cite{Teitelboim 1972}\cite{Teitelboim thesis} (DKT) in their development of covariant Hamiltonian methods. Drawing on these methods, we foliate a fixed Riemannian spacetime by a sequence of spacelike hypersurfaces; these hypersurfaces being the proper relativistic generalization of an instant to curved spacetimes. By decomposing spacetime into a succession of spatial slices in this manner, we replace the local Lorentz symmetry of spacetime with the requirement of \emph{foliation invariance}, which is, in turn, implemented by the notion of \emph{path independence}, in the manner of Teitelboim and Kucha\v{r}: if an initial and final instant can be reached by many intermediate paths, then all such paths should agree.

The application of this scheme to ED is as follows. We proceed first by modifying the notion of an instant in ED --- normally defined solely by the state $\rho$ and $\phi$ --- to be, in addition, labeled by a curved spacelike hypersurface. From here, the evolution of the statistical state $\rho$ and $\phi$ from one curved surface to the next is achieved through a \emph{local} time updating scheme, which introduces a set of local time parameters, one per spatial point; this updates the variables $\rho$ and $\phi$ in a local fashion.\footnote{This notion of local time in ED should be taken to be synonymous with the ``many-time" moniker of Dirac, as well as the ``many-fingered time" of Wheeler.} A covariant dynamics is then obtained by requiring the joint dynamics of $\rho$ and $\phi$ to be path independent, \emph{\`{a} la} Teitelboim and Kucha\v{r}.

The conditions for implementing this principle are a set of brackets for the generators of local deformations of the hypersurfaces and epistemic state. These generators are required to close in a manner that is identical to that of the algebra of hypersurface deformations; this particular algebra itself being derived from the purely kinematic criterion of \emph{embeddability}. However, whereas DKT attempt to satisfy this algebra by introducing local Hamiltonian generators and Poisson brackets, here we pursue a strategy where the algebra is a set of Lie brackets among generators that are given by a set of vector field flows, i.e. derivative operators, acting on the state space of $\rho$ and $\phi$. The result, which we believe to be novel, is that the requirement of path independence \textit{necessarily} endows the variables $\rho$ and $\phi$ with all the properties of a canonical pair. What's more is that, as mentioned above, the algebra of path independence is sufficiently restrictive so as to single out the quantum potential, and thus quantum dynamics, among all the possible dynamical models.

In the current work, the model of covariant ED sketched above is used for constructing a theory for a single quantum scalar field $\chi\left(x\right)$ in a background Riemannian spacetime. This work is a natural extension of \cite{Ipek/Abedi/Caticha 2017}, in which the author, as well as M. Abedi and A. Caticha, develop a similar scheme --- the adoption of path independence as a criterion for updating --- but achieves this by way of adopting structures, such as Poisson brackets and Hamiltonian generators; as was originally done by DKT. The current work builds on these developments by discarding the necessity of assuming the canonical framework and, moreover, by deriving the quantum potential to boot; something which was possible, but not present in that work.

The structure of this work is as follows. Section \ref{Section 2} introduces the subject matter as well as the relevant information for performing inference. Section \ref{Section 3} defines entropic time and our local-time scheme for updating. Section \ref{Section 4} introduces the kinematics of spacetime embeddings. Section \ref{Section 5} introduces Teitelboim's argument for path independence and is involved with satisfying the resulting Lie bracket constraints in ED, while in section \ref{Section 6} we discuss the analogous formulation within the canonical framework. The implications of our work and possibilities for future work are explored in the conclusion, section \ref{Section 7}.
\section{Entropic dynamics}
\label{Section 2}
The initial step in any sort of statistically modeling is to define clearly the subject matter: the choice of microstates about which we are making inferences. Here we consider a scalar field $\chi \left( x\right) $, which we will often denote as $\chi \left( x\right) =\chi _{x}$. Unlike the standard Copenhagen interpretation of quantum theory where observables have definite values only when elicited through an experiment, in the ED approach these fields have definite values at all times. And, although the field values are definite, they are uncertain; and this uncertainty will be quantified by a probability distribution. 

The field configurations $\chi_{x} $ live on a $3$-dimensional curved surface $\sigma $, the points of which are labeled by coordinates $x^{i}$ ($i=1,2,3$). The space $\sigma $ is endowed with a metric $g_{ij}$ induced on it by the non-dynamical background space-time in which $\sigma $ is embedded; thus $\sigma $ is an embedded hypersurface, which for simplicity we shall refer simply as a \textquotedblleft surface\textquotedblright . The field $\chi _{x}$ is, by definition, a scalar with respect to the  three-dimensional diffeomorphisms of the surface $\sigma $.

The $\infty $-dimensional space of all possible field configurations is the configuration space $\mathcal{C}$. A single field configuration, labelled $\chi $, is represented by a point in this space $\chi\in\mathcal{C}$. It is convenient then to model our state of incomplete information about the field $\chi$ by probabilities $P \lbrack \chi ]$ distributed over the configuration space $\mathcal{C}$.\footnote{Our notation follows the standard conventions of the field. That is, ``functions" or functionals $f:\mathcal{C}\rightarrow\mathbb{R}$ are given square brackets $f=f[\chi]$. A variation of a functional $f[\chi]$ is given by \[\delta f[\chi]=\int dx \frac{\delta f[\chi]}{\delta\chi_{x}} \, \delta\chi_{x},\] where the coefficient of $\delta\chi_{x}$ defines the \emph{functional} derivative, and where we have introduced the notation $d^{3}x\equiv dx$ for simplicity. Finally, integration over the domain $\mathcal{C}$ includes the notion of a \emph{functional} integration measure denoted by $D\chi\sim \prod_{x}d\chi_{x}$. A good review of the functional calculus of fields is given in ch. 2 of \cite{Greiner Field Quant}.}
\paragraph*{The Maximum Entropy scheme}
Our goal here is to predict the dynamical behavior of the scalar field $\chi$. The state of knowledge we wish to express should account, not only for our uncertainty in the field $\chi$, but also our uncertainty in what subsequent field values $\chi^{\prime}=\chi+\Delta\chi$ we ought to expect.

At this stage of the affairs, our state of knowledge with respect to these expected future steps $\Delta\chi$ of the field $\chi$ is maximally ignorant; we are completely uninformed about the behavior of $\chi$. The probability distribution $Q[\chi^{\prime}|\chi]$ that encodes this information is one that neglects correlations between degrees of freedom
\begin{equation}
Q[\chi^{\prime}|\chi]\sim\prod_{x} Q(\chi^{\prime}_{x}|\chi_{x}),
\end{equation}
and is also completely \emph{unconstrained} in the possible step sizes $\Delta\chi$. That is, the probability $Q[\chi^{\prime}|\chi]$ that reflects this state of complete ignorance is given by a \emph{uniform} distribution $Q[\chi^{\prime}|\chi]=Q=\text{ constant.}$\footnote{Uniform distributions are useful for conveying state of complete ignorance, but are known to be non-normalizable and thus mathematically undesirable. One work-around to this issue is for $Q[\chi^{\prime}|\chi]$ to be a Gaussian of sufficiently broad width; for short steps $\Delta\chi$ the transition probability is unaffected by this choice.}

Clearly such an uninformative informational state is insufficient for making accurate predictions of physical phenomenon --- to make reasonable inferences, more information is needed. And, when such information is, indeed, made available, our next step should be to assimilate this information into our inferences. This is accomplished by way of \emph{updating} the distribution $Q[\chi^{\prime}|\chi]$ --- now called a prior distribution, or just \emph{prior} --- to a \emph{posterior} $P[\chi^{\prime}|\chi]$ that reflects this new information.

With regards to choosing the appropriate posterior, there exists, a \emph{universal} method for updating probabilities when such new information is supplied, this is the Maximum Entropy (ME) method \cite{Caticha 2012}\cite{Giffin}. The method can be presented algorithmically as follows:
\begin{itemize}
\item[1.] Establish the prior $Q[\chi^{\prime}|\chi]$ that requires updating.
\item[2.] Obtain that information which is deemed relevant for updating: this obliges us to identify the family of posteriors $\left\{P[\chi^{\prime}|\chi]\right\}$ that are consistent with this information
\begin{equation*}
Q[\chi^{\prime}|\chi]\quad \overset{\text{new information}}{\xrightarrow{\hspace*{2.5cm}}}\quad P[\chi^{\prime}|\chi].
\end{equation*}
\item[3.] Define a relative entropy $S[P,Q]$ for ranking the posteriors $P[\chi^{\prime}|\chi]$ relative to the prior $Q[\chi^{\prime}|\chi]$.
\item[4.] Select the optimal $P[\chi^{\prime}|\chi]$; the one that \emph{maximizes} $S[P,Q]$ subject to the constraints.
\end{itemize}
\paragraph*{Constraints} New information is supplied in the form of constraints. We have proceeded thus far without imposing any restrictions on the dynamics of the field $\chi$. To this end, our first constraint involves altering the microscopic model of the field: in ED, motion is assumed to be continuous, such that large changes follow from the accumulation of many infinitesimally small steps. As a consequence, we require knowledge only of the infinitesimal steps $\chi$ to $\chi^{\prime}=\chi+\Delta\chi$, as finite changes are determined from the iteration of many small ones.

In practice, the requirement of continuity is applied by imposing, at every point $x$, the constraint
\begin{equation}
\left\langle \left( \Delta \chi _{x}\right) ^{2}\right\rangle \equiv \int D\chi ^{\prime }\,P\left[ \chi ^{\prime }|\chi\right] \left( \Delta \chi _{x}\right) ^{2}=\kappa _{x}
\label{Constraint 1}
\end{equation}%
on the posterior $P[\chi^{\prime}|\chi]$. The $\kappa _{x}$ are required to be small quantities, and in the limit $\kappa _{x}\to 0$, will ensure that the motion is continuous. On its own, the constraints (\ref{Constraint 1}) lead, however, to a dynamics that is not sufficiently rich: the degrees of freedom do not demonstrate any correlations. A more interesting dynamics is produced by imposing an additional \textit{global} constraint
\begin{equation}
\left\langle \Delta \phi\right\rangle =\int D\chi ^{\prime }\,\int dx\,\,P\left[ \chi ^{\prime }|\chi
\right]  \,\Delta \chi _{x}%
\frac{\delta \phi \left[ \chi \right] }{\delta \chi _{x}}=\kappa ^{\prime },
\label{Constraint 2}
\end{equation}%
where $\kappa^{\prime}\to 0$ is also a small quantity. This constraint serves to couple and induce correlations among the degrees of freedom, and leads, eventually, to properties such as entanglement and interference that are typical of quantum theory. These properties are characterized by the functional $\phi\left[\chi\right]$, which is called the \textit{drift potential}, and which plays the role of a pilot wave that guides the field; it will later be related to the phase of the quantum state $\Psi$.\footnote{In alternative formulations of ED, the role of the pilot wave is not played by a drift potential, but by the entropy of some auxillary ``$y$-variables." Here this distinction does not concern us.}\footnote{Before proceeding, please note that since $\chi$ and $\Delta\chi$ transform as scalars on $\sigma$, the constraint (\ref{Constraint 2}) is also coordinate invariant (under surface diffeomorphisms) only if the functional derivative $\delta/\delta\chi_{x}$ transforms as a scalar density.}
\paragraph*{The transition probability}
To select a particular posterior $P[\chi^{\prime}|\chi]$ that satisfies the constraints (\ref{Constraint 1}) and (\ref{Constraint 2}), we first define the relative entropy $S[P,Q]$
\begin{equation}
S\left[ P,Q \right] =-\int
D\chi ^{\prime }P\left[ \chi ^{\prime }|\chi\right]  \log \frac{P\left[ \chi ^{\prime }|\chi\right]  }{Q\left[ \chi ^{\prime }|\chi\right]  },
\label{Entropy Transition Probability}
\end{equation}%
for the purposes of ranking the posteriors relative to the prior. The most preferred distribution $P\left[ \chi ^{\prime }|\chi\right]$ is the one that maximizes $S[P,Q]$. The result of performing this maximization subject to the constraints (\ref{Constraint 1}), (\ref{Constraint 2}), and normalization, leads to a Gaussian transition probability distribution
\begin{equation}
P\left[ \chi ^{\prime }|\chi\right] =\frac{1}{Z%
\left[ \alpha _{x},\alpha^{\prime},g_{x}\right] }\,\exp -\frac{1}{2}\int dx\,g_{x}^{1/2}\alpha _{x}\left( \Delta \chi _{x}-\frac{\alpha^{\prime}}{g_{x}^{1/2}\alpha _{x}}\frac{\delta \phi \left[ \chi \right] }{\delta \chi
_{x}}\right) ^{2}.  \label{Trans Prob}
\end{equation}%
Here $Z\left[ \alpha _{x},\alpha^{\prime},g_{x}\right] $ is a normalization constant, $\alpha _{x}$ and $\alpha^{\prime}$ are the Lagrange multipliers for the constraints (\ref{Constraint 1}) and (\ref{Constraint 2}), respectively, and the scalar density $g_{x}^{1/2}=\det|g_{ij}(x)|^{1/2}$ has been introduced so that things transform appropriately.

As is typical, the Lagrange multipliers $\alpha_{x}$ and $\alpha^{\prime}$ are chosen to enforce the constraints (\ref{Constraint 1}) and (\ref{Constraint 2}). With regards to the single multiplier $\alpha^{\prime}$ for the global constraint (\ref{Constraint 2}), this multiplier is not featured in any final results, and thus no generality is lost by simply letting $\alpha^{\prime}=1$.\footnote{A more in-depth discussion on the role and interpretation of $\alpha^{\prime}$ is given in \cite{Bartolomeo alpha prime}. The essential result is that $\alpha^{\prime}$ is important in establishing that ED includes Bohmian dynamics as a special case.} The scalar valued Lagrange multipliers $\alpha_{x}$, on the other hand, must be chosen to satisfy the constraint (\ref{Constraint 1}), with $\kappa_x\to 0$. Inspection of (\ref{Trans Prob}) shows that the probability of finite steps $\Delta\chi_{x}$ is vanishingly small only in the limit $\alpha _{x}\rightarrow \infty $, and thus $\alpha_{x}$ must be chosen accordingly. For context, in previous work \cite{Caticha Ipek 2014}, where ED was formulated in flat space, $\alpha_x$ was restricted to be a spatial constant $\alpha_{x}=\alpha$ so as to reflect the translational symmetry of flat space. Here no such restriction is imposed. In fact, relaxing the assumption of a single global $\alpha$ in favor of a non-uniform spatial scalar $\alpha_x$ will be a key element in implementing our scheme for a local entropic time.
\paragraph*{Expected motion}
The infinitesimal dynamics of the field $\chi$ can be now be determined. The Gaussian form of (\ref{Trans Prob}) allows us to present a generic
change $\Delta \chi _{x}=\left\langle \Delta \chi _{x}\right\rangle +\Delta
w_{x}$ as resulting from an expected drift $\left\langle \Delta \chi
_{x}\right\rangle $ plus fluctuations $\Delta w_{x}$. It is a simple matter to determine the moments. The expected step is%
\begin{equation}
\left\langle \Delta \chi _{x}\right\rangle =\frac{1}{\alpha_x\,g_{x}^{1/2}}%
\frac{\delta \phi \left[ \chi \right] }{\delta \chi _{x}}\equiv \Delta \bar{%
\chi}_{x},  \label{Exp Step 1}
\end{equation}%
while the fluctuations satisfy
\begin{equation}
\left\langle \Delta w_{x}\right\rangle =\left\langle \Delta \chi
_{x}-\left\langle \Delta \chi _{x}\right\rangle \right\rangle =0\hspace{.25cm}\textrm{and}\hspace{.25cm}\left\langle \Delta w_{x}\Delta w_{x^{\prime }}\right\rangle =\frac{1}{\alpha_x\,g_{x}^{1/2}}\delta _{xx^{\prime }}.\label{Sq Fluctuations 1}
\end{equation}%
We see from these that while the expected steps satisfy $\Delta \bar{\chi}_{x}\sim 1/\alpha_{x}$, the fluctuations behave as $\Delta w_{x}\sim 1/\alpha_{x}^{1/2}$, so that in the limit $\alpha_{x}\rightarrow \infty $, the fluctuations dominate the motion. The resulting trajectory is thus continuous, but not differentiable: just as in Brownian motion.

\section{Entropic time}
\label{Section 3}
In ED finite changes are generated as a succession of many small steps. To enact this process, we require a tool for iterating these small steps into finite ones and monitoring their progression. This involves introducing a notion of $\emph{time}$ into ED, which includes the designation of an instant, how those instants are ordered, and a determination of the separation between those instants. We call this scheme \textit{entropic time}.
\paragraph*{Entropic instant}
Of particular relevance to any dynamical theory is the establishment of an instant of time. In ED this involves several ingredients (1) a spacelike surface $\sigma$ that provides a criterion of simultaneity and codifies spatial relations among degrees of freedom, (2) a specification of the Cauchy data on that surface, here the statistical state, $\rho$ and $\phi$, and (3) an entropic step in which the statistical state at one instant is updated to generate the state at the next instant.

The key difference here between previous work \cite{Caticha Ipek 2014}\cite{Bartolomeo et al 2014} and the present work is the explicit inclusion of the surface $\sigma$. And indeed, while the dynamical content of interest is contained in the variables $\rho$ and $\phi$, the surface $\sigma$ plays a crucial role, not only as a kinematic tool for organizing our inferences, but also as a vehicle to express alternative notions of simultaneity; a key component of a relativistic theory.
\paragraph*{Spacetime Kinematics}
In order to discuss the scheme for updating from one instant to the next, it is useful to first introduce some spacetime kinematics. We deal here with a curved Riemannian spacetime $\mathcal{M}$, where spacetime events are labeled by the coordinates $X^{\mu}$ ($\mu=0,1,2,3$), and with spacetime metric $g_{\mu\nu}\left(X^{\alpha}\right)$. We follow Dirac in foliating $\mathcal{M}$ by a sequence of spacelike surfaces $\left\lbrace\sigma\right\rbrace$, each with coordinates $x^{i}$ ($i=1,2,3$). The spacetime coordinates $X^{\mu}$ can then be alternatively viewed as a set of four (surface) scalar \textit{embedding} variables $X^{\mu}\left(x\right)$ for the three spatial coordinates $x^{i}$; these variables are needed to establish the location of the point $x^{i}$ at each surface $\sigma$ in spacetime.

The induced metric of the surface $\sigma$ is determined from the metric of the enveloping spacetime according to the standard procedure
\begin{equation}
g_{ij}=g_{\mu\nu}\left(X^{\alpha}\right)X^{\mu}_{ix}X^{\nu}_{jx},\quad
\text{where}\quad X^{\mu}_{ix}=\frac{\partial X^{\mu}\left(x\right)}{\partial x^{i}}.
\label{Induced surface metric}
\end{equation}
The quantity $X^{\mu}_{ix}$, introduced in (\ref{Induced surface metric}), projects spacetime vectors onto the surface $\sigma$, and is defined by the relation $g_{\mu\nu} \, X^{\mu}_{i} \, n^{\nu}=0$, where $n^{\mu}$ ($n_{\mu}n^{\mu}=-1$) is the vector normal to the surface. It is important to note that although, in general, the surface metric $g_{ij}$ will vary from one surface to the next, neither the surface metric $g_{ij}$ nor the spacetime metric $g_{\mu\nu}$ are dynamical. We deal here with a field theory in a non-dynamical background spacetime.

Following Teitelboim and Kucha\v{r}, we consider an infinitesimal deformation of a surface $\sigma$ into a neighboring surface $\sigma^{\prime}$. This is given by the deformation
\begin{equation}
\delta \xi^{\mu }=\delta \xi^{\perp} \,n^{\mu }+\delta \xi^{i}X_{i}^{\mu }~.
\label{deformation vector}
\end{equation}%
The normal and tangential components of the deformation (related, respectively, to the lapse and shift functions) are collectively denoted $(\delta \xi^{\perp} ,\delta \xi^{i})=\delta \xi ^{A}$ and are given by%
\begin{equation}
\delta \xi^{\perp}_{x}=n_{\mu x}\delta X_{x}^{\mu }\quad \text{and}\quad\delta \xi_{x}^{i}=X_{\mu x}^{i}\delta X_{x}^{\mu }.
\end{equation}%
The deformation $\delta \xi^{A}$ is said to connect the point $x$ on the surface $\sigma$ with the same point $x$ on the neighboring surface $\sigma^{\prime}$. As a matter of convention, the deformations are identified by only by their components $\delta \xi^{\perp} $ and $\delta \xi^{i}$. This allows us to speak unambiguously about applying the same deformations to different surfaces; a notion that is crucial for implementing the path independence criterion.
\paragraph*{Duration}
In ED the notion of separation between instants, or duration, is defined so that motion looks simple. Since for small steps the dynamics is dominated by the fluctuations, a convenient way to define the duration is then to just make the fluctuations look as simple as possible. This is accomplished by choosing choosing the multipliers $\alpha_{x}$ in (\ref{Sq Fluctuations 1}) appropriately.

Moreover, regarding the notion of duration, notice that we deal here with instants characterized by curved surfaces in spacetime. This implies that any notion of duration ought to be \textit{local} in character to accommodate this fact. To this end, consider deforming an initial surface $\sigma$ by a deformation $\delta \xi_{x}^{\mu}$ that is everywhere normal to it. The interval that points from the point $x$ on $\sigma$ to the point $x$ on $\sigma^{\prime}$ is then just the proper time of the geodesic stretching from $\sigma$ to $\sigma^{\prime}$, given by the normal component $\delta\xi^{\perp}_{x}$ of the deformation. This gives us the local notion of duration that we seek. Implementing this notion in ED is achieved by letting
\begin{equation}
\alpha _{x}=\frac{1}{\eta \delta \xi^{\perp}_{x}}\quad \text{so that}\quad
\left\langle \Delta w_{x}\Delta w_{x^{\prime }}\right\rangle =\frac{\eta
\,\delta \xi^{\perp}_{x}}{g_{x}^{1/2}}\delta _{xx^{\prime }}~,
\label{Duration}
\end{equation}%
where $\eta $ is just a constant that relates the units of time to those of $\chi $.

With this choice of multiplier $\alpha_{x}$, the transition probability $P[\chi^{\prime}|\chi]$ in (\ref{Trans Prob}) can be interpreted as providing the likelihood that a configuration $\chi$ on a surface $\sigma$ jumps to a configuration $\chi^{\prime}$ on a neighboring surface $\sigma^{\prime}$ that is obtained from a purely normal deformation of $\sigma$, i.e. $\delta\xi^{A}_{x}=(\delta\xi^{\perp}_{x},0)$.\footnote{The inclusion of more general deformations, including tangential components, will be addressed later.} Thus the transition probability (\ref{Trans Prob}) can be said to describe a Wiener process, albeit in a non-standard form involving the notion of local times.
\paragraph*{Ordered instants}
In ED, the complete state of knowledge we have of our system expresses, not only, the uncertainty we have of $\chi$ at an instant, but also what subsequent configuration $\chi^{\prime}=\chi+\Delta\chi$ that we expect will occur. The distribution that expresses this informational state is given by the \textit{joint} probability $\rho\left[\chi^{\prime},\chi\right]=P\left[\chi^{\prime}|\chi\right]\rho\left[\chi\right]$, where $P[\chi^{\prime}|\chi]$ is the transition probability given in (\ref{Trans Prob}).

Supposing that $\rho[\chi]$ is known (say it is given as an initial condition), and given that we have an estimate $P[\chi^{\prime}|\chi]$ for what changes can be expect, we might ask, what is the probability of the subsequent configuration $\chi^{\prime}$? To obtain this probability we simply apply the ``sum rule" of probability theory to the joint distribution to obtain\footnote{See \cite{Caticha 2012}\cite{Caticha 2010a} for a more detailed exposition on entropic time.}
\begin{equation}
\rho ^{\prime }\left[ \chi ^{\prime } \right] =\int D\chi \,\rho\left[ \chi ^{\prime },\chi\right]=\int D\chi \,P\left[ \chi ^{\prime }|\chi\right] \,\rho \left[ \chi\right].
\label{Evolution equation}
\end{equation}%
Note that this step requires no assumptions, this result is guaranteed purely by the rules of probability. The assumption, however, comes in the interpretation. If $\rho \left[ \chi \right]$ is said to give the statistical state at an instant, then the statistical state at the subsequent instant is said to be given by $\rho ^{\prime }\left[ \chi ^{\prime } \right]$. Accordingly, we write  $\rho \left[ \chi \right]=\rho_{\sigma}\left[\chi\right]$ and $\rho ^{\prime }\left[ \chi ^{\prime } \right]=\rho_{\sigma^{\prime}}\left[ \chi ^{\prime } \right]$ to denote the states at the instants labeled by $\sigma$ and $\sigma^{\prime}$, respectively. The equation (\ref{Evolution equation}) can now be written as
\begin{equation}
\rho_{\sigma^{\prime}}\left[ \chi ^{\prime } \right]=\int D\chi \,P\left[ \chi ^{\prime }|\chi\right] \,\rho_{\sigma} \left[ \chi\right]. \label{Dynamical equation}
\end{equation}%
This is the basic dynamical equation that dictates the update of $\rho_{\sigma}\left[\chi\right]$ from one instant to the next.\footnote{Notice also that there is a natural arrow to the flow of entropic time. While $P\left[ \chi ^{\prime }|\chi\right]$ is obtained by maximizing an entropy, the `time-reversed' probability $P\left[ \chi |\chi^{\prime }\right]$ is only given by Bayes theorem; thus the two are distinct.} However, for the dynamics of $\rho_{\sigma}\left[\chi\right]$ to remain consistent with the deformations of the underlying surfaces $\sigma$, its evolution must itself satisfy an additional criterion that is typical of surfaces in spacetime: \textit{path independence}: if there are many ways to proceed from an initial surface to a final surface, then all ways should agree. Details on the implementation of this are the subject of subsequent sections. 
\paragraph{The local-time diffusion equations}
The updating protocol (\ref{Evolution equation}), expressed in integral form, together with the notion of a local duration (\ref{Duration}), can be used to determine the variation of the probability $\rho_{\sigma}$ as it evolves from one surface to the next (in a purely normal fashion). This variation is given by (see derivation in Appendix \ref{appendix FP})
\begin{equation}
\delta \rho _{\sigma }\left[ \chi \right] =\int dx\frac{\delta \rho _{\sigma
}\left[ \chi \right] }{\delta \tilde{N}_{x}}\delta \xi^{\perp}_{x}=-\int
dx\frac{1}{g_{x}^{1/2}}\frac{\delta }{\delta \chi _{x}}%
\left( \rho _{\sigma }\left[ \chi \right] \frac{\delta \Phi _{\sigma }\left[
\chi \right] }{\delta \chi _{x}}\right)\delta \xi^{\perp}_{x}, \label{FP variation rho}
\end{equation}%
where we have introduced the operator $\delta/\delta\tilde{N}_{x}$, which generates the local dynamical evolution of the probability $\rho$. We have also introduced in (\ref{FP variation rho}), for convenience, the quantity
\begin{equation}
\frac{\Phi_{\sigma }\left[ \chi \right]}{\eta} =\phi _{\sigma }\left[ \chi \right]
- \log \rho _{\sigma }^{1/2}\left[ \chi \right],
\label{Definition Phi}
\end{equation}
whose interpretation will become clearer shortly. For arbitrary choices of the local separations $\delta\xi_{x}^{\perp}$ we obtain an infinite set of local equations, one for each spatial point, 
\begin{equation}
\frac{\delta \rho _{\sigma }}{\delta \tilde{N}_{x}}=-\frac{1}{g_{x}^{1/2}}\frac{\delta }{\delta \chi _{x}}\left( \rho _{\sigma }\,\frac{\delta \Phi _{\sigma}}{\delta \chi _{x}}\right).
\label{Local Time FP equation}
\end{equation}%

To see that this set of local equations is the straightforward generalization of dynamics to curved spacetime, consider the special case when both surfaces $\sigma $ and $\sigma ^{\prime }$
happen to be flat. It follows then that $g_{x}^{1/2}=1$ and $\delta \xi^{\perp}_{x}=\delta t$ are constants, so that eq.(\ref{FP variation rho}) leads to 
\begin{equation}
\frac{\partial \rho _{t}\left[ \chi \right] }{\partial t}=-\int dx\frac{%
\delta }{\delta \chi _{x}}\left( \rho _{t}\left[ \chi \right] \frac{\delta
\Phi _{t}\left[ \chi \right] }{\delta \chi _{x}}\right),
\label{FP c}
\end{equation}%
which we recognize as a standard diffusion, or Fokker-Planck, equation written as a continuity equation for the flow of probability in configuration space $\mathcal{C}$.\footnote{A result that agrees with the Fokker-Planck equation in \cite{Caticha 2013}\cite{Caticha Ipek 2014} for the ED of quantum relativistic scalar fields in a flat spacetime.} Accordingly, we will refer to the set of local equations given by (\ref{Local Time FP equation}) as the \textquotedblleft local-time Fokker-Planck\textquotedblright\ equations (LTFP). These equations describe the flow of probability with a current velocity $V_{x}\left[ \chi \right] =\delta \Phi /\delta \chi _{x}$\thinspace . Eventually, the functional $\Phi $ will be identified as the Hamilton-Jacobi functional, or the phase of the wave functional in the quantum theory. From this point onward it will be more convenient to work with the phase $\Phi$ rather than the drift potential $\phi$, but this is, of course, just a matter of preference; the results do not depend upon this choice.

\section{The Kinematics of Surface Deformations}
\label{Section 4}
The embedding of surfaces into a common spacetime is a kinematic criterion --- called ``embeddability" --- that can be studied independently of any dynamics. In turn, however, any dynamics set on these surfaces should mirror the kinematics of surface embeddings in spacetime. Here we follow the work of Teitelboim and Kucha\v{r} in developing the kinematics of surface deformations.
\subsection*{The algebra of deformations}
To study the structure of surface embeddings, consider a generic functional $T[X(x)]$ of the surface variables $X^{\mu}(x)$. A slight change in these variables $X^{\mu}\to X^{\mu}+\delta \xi^{\mu}$ induces a change in $T[X(x)]$ such that
\begin{equation}
\delta T[X(x)]=\int dx \, \delta \xi^{\mu}\frac{\delta T[X(x)]}{\delta X^{\mu}_{x}},
\label{delta T coordinate}
\end{equation}
where the variational derivatives $\delta/\delta X^{\mu}_{x}=\delta/\delta X^{\mu}(x)$ can be viewed as a set of basis vectors that form a coordinate basis in the space of surface embeddings \cite{Kuchar Hyperspace I}. More useful yet, however, is the decomposition of $\delta T$ in the directions normal and tangential to the surface
\begin{equation}
\delta T=\int dx \, \delta \xi_{x}^{A}\frac{\delta T}{\delta N_{x}^{A}}=\int dx \left(\delta \xi_{x}^{\perp}\, \frac{\delta }{\delta N_{x}}+\delta \xi_{x}^{i} \, \frac{\delta}{\delta N_{x}^{i}}\right)T,
\label{delta T coordinate}
\end{equation}
where
\begin{equation}
\frac{\delta}{\delta N_{x}}=n_{x}^{\mu}\frac{\delta}{\delta X_{x}^{\mu}}\quad\textrm{and}\quad \frac{\delta}{\delta N_{x}^{i}}=X_{i}^{\mu}\frac{\delta}{\delta X_{x}^{\mu}}
\label{Surface Generators}
\end{equation}
generate normal and tangential deformations of the surface, respectively. The differential operators $\delta/\delta N^{A}$ serve as a basis for the system of coordinates defined normal and tangential to the surfaces. Unlike, however, the derivatives $\delta/\delta X^{\mu}_{x}$ which form a coordinate basis, and thus all commute
 \[\left[\frac{\delta}{\delta X^{\mu}_{x}},\frac{\delta}{\delta X^{\nu}_{x^{\prime}}}\right]=0,\]
the operators $\delta/\delta N^{A}$ form a non-holonomic basis, and are thus defined by non-vanishing commutation relations
\begin{equation}
\frac{\delta}{\delta N^{A}_{x}}\frac{\delta}{\delta N^{B}_{x^{\prime}}}-\frac{\delta}{\delta N^{B}_{x^{\prime}}}\frac{\delta}{\delta N^{A}_{x}}=\int dx^{\prime\prime} \, \kappa^{C}_{AB}\left(x^{\prime\prime};x,x^{\prime}\right)\frac{\delta}{\delta N^{C}_{x^{\prime\prime}}},
\label{Non commuting generators}
\end{equation}
where the $\kappa^{C}_{AB}\left(x^{\prime\prime};x,x^{\prime}\right)$ are the structure ``constants" of the ``group" of deformations.\footnote{Strictly speaking, the ``group" of deformations is not a group. Indeed, while successive deformations can be composed to form another, this composition depends on the original surface that is acted upon. This manifests itself in the fact that the structure ``constants" of deformations are not constant, they depend upon the metric $g_{ij}$ of the surface.} Simply translated: deformations do not commute.
\subsection*{Embeddability}
The structure ``constants" $\kappa^{C}_{AB}\left(x^{\prime\prime};x,x^{\prime}\right)$ can be calculated by a purely kinematic argument --- embeddability --- an idea that is key for implementing the notion of dynamical path independence. Consider a surface $\sigma$ that is deformed into a surface $\sigma_{1}$ by the deformation $\delta \xi$, which is then itself deformed into a surface $\sigma^{\prime}$ by the deformation $\delta \eta$, generating the succession $\sigma\xrightarrow{\delta \xi}\sigma_{1}\xrightarrow{\delta \eta}\sigma^{\prime}$. An alternative path is given by reversing the order of the deformations: $\sigma\xrightarrow{\delta \eta}\sigma_{2}\xrightarrow{\delta \xi}\sigma^{\prime\prime}$ to obtain a final surface $\sigma^{\prime\prime}$ that, in general, differs from $\sigma^{\prime}$.

At this point, we require embeddability: for $\sigma^{\prime}$ and $\sigma^{\prime\prime}$ to be embedded in the same spacetime, there must exist a third deformation $\delta \zeta$ that connects them, such that $\sigma^{\prime}\xrightarrow{\delta \zeta}\sigma^{\prime\prime}$. The ``compensating" deformation $\delta \zeta$ that relates the final surfaces can be derived from a purely geometric argument (see \cite{Teitelboim 1972}\cite{Teitelboim thesis} for details) and has the form
\begin{equation}
\delta\zeta^{C}_{x^{\prime\prime}}=\int dx \, dx^{\prime} \kappa^{C}_{AB}\left(x^{\prime\prime};x,x^{\prime}\right)\delta\xi^{A}_{x}\delta\eta^{B}_{x^{\prime}}.
\label{Compensating deformation}
\end{equation}
The ``constants" $\kappa^{C}_{AB}\left(x^{\prime\prime};x,x^{\prime}\right)$ in (\ref{Compensating deformation}) are exactly those of (\ref{Non commuting generators}), which were used there to define the non-commutativity of deformations.\footnote{To keep our argument concise, we defer the derivation of these ``constants" to \cite{Teitelboim 1972}\cite{Teitelboim thesis}.} The result is that the generators $\delta/\delta N_{x}^{A}$ close as
\begin{subequations}
\begin{align}
[\frac{\delta}{\delta N_{x}},\frac{\delta}{\delta N_{x^{\prime}}}]&=-\left(g^{ij}_{x}\frac{\delta}{\delta N^{j}_{x}}+g^{ij}_{x^{\prime}}\frac{\delta}{\delta N^{j}_{x^{\prime}}}\right)\partial_{ix}\delta(x,x^{\prime})\label{Surface Deformations perp-perp}\\
[\frac{\delta}{\delta N^{i}_{x}},\frac{\delta}{\delta N_{x^{\prime}}}]&=-\frac{\delta}{\delta N_{x}}\,\partial_{ix}\delta(x,x^{\prime})\label{Surface Deformations tan-perp}\\
[\frac{\delta}{\delta N^{i}_{x}},\frac{\delta}{\delta N^{j}_{x^{\prime}}}]&=-\frac{\delta}{\delta N^{i}_{x^{\prime}}}\partial_{jx}\delta(x,x^{\prime})-\frac{\delta}{\delta N^{j}_{x}}\partial_{ix}\delta(x,x^{\prime}).\label{Surface Deformations tan-tan}
\end{align}
\end{subequations}
In the next section we show how requiring such an algebra to also hold for the dynamical variables $\rho$ and $\phi$ leads to a non-dissipative covariant ED. 
\section{Path independent entropic dynamics}
\label{Section 5}
The dynamical system we've developed thus far determines the evolution of the probability $\rho[\chi]$ in response to an externally prescribed background geometry and drift potential $\phi[\chi]$. A dynamics of this type is inevitably a theory of diffusion --- the local-time evolution of $\rho[\chi]$ is given by a set of local diffusion equations, the LTFP eqns.(\ref{Local Time FP equation}). While this is a completely legitimate form of dynamics, it falls short of describing the non-dissipative quantum phenomenon we wish to model. The simplest way to make progress is to promote the drift potential $\phi[\chi]$, or equivalently the functional $\Phi[\chi]$, to the level of a dynamical degree of freedom. This amounts to modifying the constraint (\ref{Constraint 2}) in response to the update of $\rho[\chi]$.

The natural question then becomes, what rule determines the precise manner in which $\Phi[\chi]$ is to be updated? Here we propose that the update of $\Phi[\chi]$ be determined, not by conserving a global energy (as in \cite{Caticha Ipek 2014}\cite{Bartolomeo et al 2014}), but by requiring the joint dynamics of $\rho[\chi]$ and $\Phi[\chi]$ to be independent of the path: the evolution from an initial state to a final state should be independent of the intermediate path. The exact manner in which this is to be implemented is the topic of the current section.\footnote{For completeness, we mention here the LTFP eqns.(\ref{Local Time FP equation}) on their own do not automatically give a covariant dynamics either. For this, one must impose path independence on the $\rho$'s alone. Indeed, it is possible to do this, but this line of research is not developed here because it is unclear what physical interpretation, if any, we could assign to this model.} 
\subsection*{Defining the formalism}
It will be convenient to represent the joint state $(\rho,\Phi)$ as being a point in a space $\Gamma$ of possible states.\footnote{Note that this is an infinitely dimensional space of greater magnitude than is typical. It is not a space of functions (such as $\mathcal{C}$) but a space of \emph{functionals}. However, as is typical, we are only semi-rigourous about the construction of this space. That is, we assume that the structures we introduce have the properties needed to support our arguments.} Thus, $\rho$ and $\Phi$ act as coordinates for $\Gamma$ such that $\gamma\equiv(\rho,\Phi)\in\Gamma$ is a point in that space. Furthermore, as we've mentioned before, we deal here with a background spacetime that has been foliated by a sequence of surfaces $\{\sigma\}$. To locate the point $x$ on each surface requires the use of four embedding, or surface, variables $X^{\mu}(x)=X^{\mu}_{x}$. The space of surface variables $X^{\mu}_{x}$ then itself forms a space of infinite dimension, called ``Hyperspace" by Kucha\v{r} \cite{Kuchar Hyperspace I}, and which we denote as $\Sigma$.

A sequence of surfaces in a foliation amount to defining a curve in $\Sigma$. To each surface $\sigma$ in this foliation we attach a statistical state $\gamma_{\sigma}=(\rho_{\sigma},\Phi_{\sigma})$ associated with that surface; this is, of course, the Cauchy data for that surface. Then, just as we have a curve $\{\sigma\}$ in $\Sigma$, there is a corresponding curve $\{\gamma\}$ in $\Gamma$ for the evolution of the states $\rho$ and $\Phi$ (Note that the curve $\{\gamma\}$ and the curve $\{\sigma\}$ have distinct origins: the former is a genuine dynamical path while the latter is a matter of kinematic choice). Since the foliation is monoparametric, each surface can be labeled by a single parameter, call it $\tau$. This label can then be used to parameterize curves in $\Sigma$, or similarly, can also be used to label the curves in $\Gamma$.

\paragraph*{Surface Generators} Once the notion of a curve has been established, a useful concept is that of a vector tangent to it. For the curve $\{\sigma\}$ in $\Sigma$, parameterized by $\tau$, this tangent vector can be written as
\begin{equation}
\frac{d}{d\tau}=\int dx \, N^{\mu}_{x}\frac{\delta}{\delta X^{\mu}_{x}}\quad\text{where}\quad N^{\mu}_{x}=\frac{\partial X^{\mu}_{x}}{\partial \tau}.
\end{equation}
The differential operators $\delta/\delta X^{\mu}_{x}$ serve as a coordinate basis for $\Sigma$, while the $N^{\mu}_{x}$ behave as components of a spacetime vector field.

From the coordinate basis, given by the $\delta/\delta X^{\mu}_{x}$, we can pass to a basis that is better suited to our purpose. The idea is analogous to the one introduced above: we perform an orthogonal decomposition of the basis vectors $\delta/\delta X^{\mu}_{x}$, given by the new basis vectors
\begin{equation}
\frac{\delta}{\delta N^{A}_{x}}=(\frac{\delta}{\delta N_{x}},\frac{\delta}{\delta N^{i}_{x}}),
\end{equation}
which are the same as those introduced in (\ref{Surface Generators}). With this, the tangent vector $d/d\tau$ becomes
\begin{equation}
\frac{d}{d\tau}=\int dx \, \left(N_{x}\frac{\delta}{\delta N_{x}}+N^{i}_{x}\frac{\delta}{\delta N^{i}_{x}}\right),
\label{Curve tau Sigma}
\end{equation}
where
\begin{equation}
N_{x}=n_{\mu x}N^{\mu}_{x}\quad\text{and}\quad N^{i}_{x}=X^{i}_{\mu x}N^{\mu}_{x}.
\end{equation}
The scalar function $N_{x}$ is called the ``lapse" function while the vector $N^{i}_{x}$ is called the ``shift"; this then further implies that $\delta/\delta N_{x}$ is a scalar density, while $\delta/\delta N^{i}_{x}$ is a covariant vector density.

The lapse and shift functions are related to the infinitesimal deformations $\delta\xi_{x}^{A}$ by the relations
\begin{equation}
\delta\xi_{x}^{\perp}=N_{x}\delta \tau\quad\text{and}\quad\delta\xi_{x}^{i}=N_{x}^{i}\delta \tau.
\end{equation}
Thus the interpretation of the lapse can also be given in terms of proper time, as was done for $\delta\xi_{x}^{\perp}$, while the shift vector $N^{i}$, on the other hand, reflects the tangential displacement inherent in passing from $\sigma_{\tau}$ to $\sigma_{\tau+\delta\tau}$. The ``local" vectors $\delta/\delta N^{A}_{x}$ that pair with the lapse and shift in (\ref{Curve tau Sigma}) are particularly important: they are the generators of local surface deformations that we introduced originally in (\ref{Surface Generators}). These are the objects that will be crucial for implementing path independence.

\paragraph*{Ensemble Generators} Similar considerations are relevant to the curve on $\Gamma$. The vector tangent to the curve parameterized by $\tau$ in $\Gamma$ is given by 
\begin{equation}
\frac{d}{d\tilde{\tau}}=\int D\chi \left(\partial_{\tau}\rho[\chi]\frac{\tilde{\delta}}{\tilde{\delta}\rho[\chi]}+\partial_{\tau}\Phi[\chi]\frac{\tilde{\delta}}{\tilde{\delta}\Phi[\chi]}\right)
\end{equation}
in the basis given by the coordinates $(\rho,\Phi)$. We have used the notation where derivatives with respect to functionals get a tilde. Also, in the following we will use the practice of denoting $\rho[\chi]=\rho_{\chi}$ or completely suppressing the $\chi$ index altogether when needed.

To make progress, we must pay special attention to the quantities $\partial_{\tau}\rho$ and $\partial_{\tau}\Phi$. For this, consider the change
\begin{equation}
\delta\rho[\sigma_{\tau}]=\rho[\sigma_{\tau+\delta\tau}]-\rho[\sigma_{\tau}],
\end{equation}
as $\rho$ is dragged from one surface to a neighboring surface (an analogous argument will hold for $\Phi$ as well). As we saw above, the passage from one surface to the next can be decomposed into motions parallel to the surface and those that are orthogonal to it. A similar decomposition can be performed here as well, leading to
\begin{equation}
\delta\rho[\sigma_{\tau}]=\int dx \, \left(N_{x}\frac{\delta\rho[\sigma_{\tau}]}{\delta \tilde{N}_{x}}+N_{x}^{i}\frac{\delta\rho[\sigma_{\tau}]}{\delta \tilde{N}^{i}_{x}}\right)\delta\tau,
\end{equation}
so that we obtain
\begin{equation}
\partial_{\tau}\rho[\sigma_{\tau}]=\int dx \, \left(N_{x}\frac{\delta\rho[\sigma_{\tau}]}{\delta \tilde{N}_{x}}+N_{x}^{i}\frac{\delta\rho[\sigma_{\tau}]}{\delta \tilde{N}^{i}_{x}}\right),
\end{equation}
where $N_{x}$ and $N^{i}_{x}$ are, respectively, the lapse and shift introduced above, and where
\begin{equation}
\frac{\delta\rho[\sigma_{\tau}]}{\delta\tilde{N}_{x}^{A}}=\left(\frac{\delta\rho[\sigma_{\tau}]}{\delta\tilde{N}_{x}},\frac{\delta\rho[\sigma_{\tau}]}{\delta\tilde{N}_{x}^{i}}\right)
\end{equation}
determines the response of the functional $\rho$ to normal and tangential deformations, respectively.

Performing a similar operation for $\Phi$, we then have that the vector tangent to the curve $\tau$ becomes
\begin{align}
\frac{d}{d\tilde{\tau}}&=\int D\chi \int dx \left(N_{x}\left(\frac{\delta\rho}{\delta \tilde{N}_{x}}\frac{\tilde{\delta}}{\tilde{\delta}\rho}+\frac{\delta\Phi}{\delta \tilde{N}_{x}}\frac{\tilde{\delta}}{\tilde{\delta}\Phi}\right)\right.\notag\\
&\hspace{2.1 cm}+\left. N^{i}_{x}\left(\frac{\delta\rho}{\delta \tilde{N}^{i}_{x}}\frac{\tilde{\delta}}{\tilde{\delta}\rho}+\frac{\delta\Phi}{\delta \tilde{N}^{i}_{x}}\frac{\tilde{\delta}}{\tilde{\delta}\Phi}\right)\right).
\end{align}
The quantities $\delta\rho/\delta \tilde{N}^{A}_{x}$ and $\delta\Phi/\delta \tilde{N}^{A}_{x}$ will be discussed shortly.

As we did above, we can take the coefficient of the lapse function $N_{x}$ to be the normal evolution generator for $\rho$ and $\Phi$, while the coefficient of the shift $N^{i}_{x}$ is the tangential generator. More explicitly we have
\begin{equation}
\frac{\delta}{\delta\tilde{N}_{x}}=\int D\chi \left(\frac{\delta\rho}{\delta \tilde{N}_{x}}\frac{\tilde{\delta}}{\tilde{\delta}\rho}+\frac{\delta\Phi}{\delta \tilde{N}_{x}}\frac{\tilde{\delta}}{\tilde{\delta}\Phi}\right),
\label{Ensemble Normal Generators}
\end{equation}
and
\begin{equation}
\frac{\delta}{\delta\tilde{N}^{i}_{x}}=\int D\chi \left(\frac{\delta\rho}{\delta \tilde{N}^{i}_{x}}\frac{\tilde{\delta}}{\tilde{\delta}\rho}+\frac{\delta\Phi}{\delta \tilde{N}^{i}_{x}}\frac{\tilde{\delta}}{\tilde{\delta}\Phi}\right),
\label{Ensemble Tangential Generators}
\end{equation}
which are the normal and tangential \emph{ensemble} generators, respectively. For conciseness we refer to these generators simply as e-generators and denote them $\delta/\delta\tilde{N}_{x}^{A}$, where $A=(\perp,i)=(\text{normal},\text{ tangential})$, as usual. Note that these e-generators are very unorthodox: they are local in the spatial coordinates $x$, but include an integration over all field configurations $\chi$. Thus these operators do, in fact, generate \emph{spatially} local deformations, even if their action in the configuration space $\mathcal{C}$ is clearly non-local.

We return now to the terms $\delta\rho/\delta \tilde{N}^{A}_{x}$ and $\delta\Phi/\delta \tilde{N}^{A}_{x}$. It turns out, fortunately enough, that most of them are known, or can be computed \emph{a priori}. That is, for example,
\begin{equation}
\frac{\delta\rho}{\delta \tilde{N}_{x}}=-\frac{1}{g_{x}^{1/2}}\frac{\delta}{\delta\chi_{x}}\left(\rho\frac{\delta\Phi}{\delta\chi_{x}}\right)
\end{equation}
is just the FP equation. Moreover, the tangential pieces can be computed straightforwardly by performing Lie derivatives along the surface
\begin{equation}
\frac{\delta\rho}{\delta \tilde{N}^{i}_{x}}=\frac{\delta\rho}{\delta\chi_{x}}\partial_{ix}\chi_{x}\quad\text{and}\quad\frac{\delta\Phi}{\delta \tilde{N}^{i}_{x}}=\frac{\delta\rho}{\delta\chi_{x}}\partial_{ix}\chi_{x}.
\end{equation}
The one unknown element is the normal evolution of $\Phi$. For this we assume at the outset a completely general form
\begin{equation}
\frac{\delta\Phi}{\delta \tilde{N}^{i}_{x}}=G[\rho,\Phi;\chi,X^{\mu}].
\end{equation}
The idea is to determine the allowed forms of the functional $G[\rho,\Phi;\chi,X^{\mu}]$ by requiring that the joint evolution of $\rho$ and $\Phi$, together with the surfaces $\sigma$, evolve consistently in spacetime.
\subsection*{Path independent dynamics}
The criterion of embeddability is purely kinematic in nature. Dynamics is introduced into the picture by requiring that the generators of the dynamics form an algebra that mirrors that of the ``group" of deformations. The basic idea is path independence: if we evolve the data from an initial surface to a final one through various intermediate foliations, our results should not depend on our choice of the intervening path. This condition, consequently, implements a foliation invariant evolution, which in turn reflects the local relativistic invariance we expect of a covariant theory.

To execute these notions, we first adjoin the space of surfaces $\Sigma$ with the ensemble state space $\Gamma$, such that we have $\mathcal{T}=\Sigma\,\otimes\,\Gamma$. Now, consider a functional $F_{\sigma}[\rho,\Phi;X^{\mu}]$ of the embedding variables and statistical state, denoted $F_{\sigma}$ for convenience; the value of $F_{\sigma}$ is defined unambiguously by the surface $\sigma$ that it resides on. The variation in $F_{\sigma}$ as $\sigma$ is deformed into $\sigma^{\prime}$ by a deformation $\delta\xi$ is given by
\begin{equation}
\delta F_{\sigma}=\int dx \, \frac{\delta F_{\sigma}}{\delta N_{x\text{T}}^{A}}\delta\xi^{A}_{x},
\end{equation}
where we have introduced the total generator
\begin{equation}
\frac{\delta}{\delta N^{A}_{x\text{T}}}=\frac{\delta}{\delta N^{A}_{x}}+\frac{\delta}{\delta \tilde{N}^{A}_{x}},
\label{Total Generator}
\end{equation}
which generates deformations for the surfaces as well as $\rho$ and $\Phi$.

In a construction that is analogous to that used for embeddability, consider now the evolution of $F_{\sigma}$ along the sequence of surfaces $\sigma\xrightarrow{\delta \eta}\sigma_{2}\xrightarrow{\delta \xi}\sigma^{\prime\prime}$ obtained by application of the deformation $\delta\eta$ then $\delta\xi$. Reversing now the order of those deformations and applying the compensating deformation $\delta\zeta$ we can also evolve $F_{\sigma}$ along the path $\sigma\xrightarrow{\delta \xi}\sigma_{1}\xrightarrow{\delta \eta}\sigma^{\prime}\xrightarrow{\delta\zeta}\sigma^{\prime\prime}$. The notion of path independence is then implemented by requiring consistency: the state of the system on the final surface $\sigma^{\prime\prime}$ should be the same regardless of the path taken. This is imposed mathematically by requiring that
\begin{equation}
F\left[\sigma\to\sigma_{1}\to\sigma^{\prime}\to\sigma^{\prime\prime}\right] =F\left[\sigma\to\sigma_{2}\to\sigma^{\prime\prime}\right],
\label{Consistency condition}
\end{equation}
which for arbitrary deformations $\delta\xi$ and $\delta\eta$ is equivalent to
\begin{equation}
\left( \frac{\delta}{\delta N^{B}_{x^{\prime}\text{T}}} \, \frac{\delta}{\delta N^{A}_{x\text{T}}} - \frac{\delta}{\delta N^{A}_{x\text{T}}} \, \frac{\delta}{\delta N^{B}_{x^{\prime}\text{T}}} \, \right)F_{\sigma}=\int dx^{\prime\prime} \, \kappa^{C}_{AB}(x^{\prime\prime};x,x^{\prime}) \, \frac{\delta}{\delta N^{C}_{x^{\prime\prime}\text{T}}}F_{\sigma}.
\label{Path Independence F}
\end{equation}
Moreover, since this relation is linear in $F_{\sigma}$, which is itself an arbitrary functional, it must be the case that the generators themselves must satisfy the relations
\begin{equation}
\frac{\delta}{\delta N^{B}_{x^{\prime}\text{T}}} \, \frac{\delta}{\delta N^{A}_{x\text{T}}} - \frac{\delta}{\delta N^{A}_{x\text{T}}} \, \frac{\delta}{\delta N^{B}_{x^{\prime}\text{T}}} \, =\int dx^{\prime\prime} \, \kappa^{C}_{AB}(x^{\prime\prime};x,x^{\prime}) \, \frac{\delta}{\delta N^{C}_{x^{\prime\prime}\text{T}}}.
\end{equation}
Or, in explicit terms, using the structure ``constants" $\kappa^{C}_{AB}(x^{\prime\prime};x,x^{\prime})$ that lead to eqns.(\ref{Surface Deformations perp-perp})-(\ref{Surface Deformations tan-tan}), we have
\begin{subequations}
\begin{align}
[\frac{\delta}{\delta N_{x\text{T}}},\frac{\delta}{\delta N_{x^{\prime}\text{T}}}]&=-\left(g^{ij}_{x}\frac{\delta}{\delta N^{j}_{x\text{T}}}+g^{ij}_{x^{\prime}}\frac{\delta}{\delta N^{j}_{x^{\prime}\text{T}}}\right)\partial_{ix}\delta(x,x^{\prime})\label{Total Deformations perp-perp}\\
[\frac{\delta}{\delta N^{i}_{x\text{T}}},\frac{\delta}{\delta N_{x^{\prime}\text{T}}}]&=-\frac{\delta}{\delta N_{x\text{T}}}\,\partial_{ix}\delta(x,x^{\prime})\label{Total Deformations tan-perp}\\
[\frac{\delta}{\delta N^{i}_{x\text{T}}},\frac{\delta}{\delta N^{j}_{x^{\prime}\text{T}}}]&=-\frac{\delta}{\delta N^{i}_{x^{\prime}\text{T}}}\partial_{jx}\delta(x,x^{\prime})-\frac{\delta}{\delta N^{j}_{x\text{T}}}\partial_{ix}\delta(x,x^{\prime}).\label{Total Deformations tan-tan}
\end{align}
\end{subequations}
These are the constraint conditions that must be satisfied by our dynamics. Our task is now to determine the possible solutions to these equations.
\subsection*{Satisfying the kinematical Lie brackets}
The goal now is to solve the constraint equations (\ref{Total Deformations perp-perp})-(\ref{Total Deformations tan-tan}). A convenient aspect of the decomposition of deformations into normal and tangential pieces is that it isolates the dynamically relevant pieces --- the normal deformations --- from the kinematic pieces, which are parallel to the surface. As one might guess, it is the purely dynamical Lie bracket (\ref{Total Deformations perp-perp}) that is more difficult to satisfy, and thus we delay treating it for later. Our current goal is to show how the other two Lie brackets, eqns.(\ref{Total Deformations tan-perp}) and (\ref{Total Deformations tan-tan}) can be satisfied in a straightforward fashion.
\paragraph*{Tangential-Tangential Lie bracket}
The generators of spatial surface diffeomorphisms form a genuine (infinite dimensional) Lie group with an algebra given by the Lie bracket (\ref{Total Deformations tan-tan}). To determine the elements of this group explicitly, a natural step is to separate the total generator (\ref{Total Generator}) into its surface and ensemble pieces given by (\ref{Surface Generators}) and (\ref{Ensemble Tangential Generators}), respectively. Recalling the variations
\begin{equation}
\frac{\delta\rho_{\sigma}}{\delta\tilde{N}^{i}_{x}}=\frac{\delta\rho_{\sigma}}{\delta\chi_{x}}\partial_{ix}\chi_{x}\quad\text{and}\quad\frac{\delta\Phi_{\sigma}}{\delta\tilde{N}^{i}_{x}}=\frac{\delta\Phi_{\sigma}}{\delta\chi_{x}}\partial_{ix}\chi_{x}
\label{Ensemble Tangential Derivatives}
\end{equation}
and using the tangential e-generator, given in (\ref{Ensemble Tangential Generators}), we observe that the tangential e-generator is \emph{independent} of the surface variables. On the other hand, the surface variables are themselves independent of $\rho$ and $\Phi$. As a consequence, the Lie bracket (\ref{Total Deformations tan-tan}) decomposes into pieces that are purely surface and purely ensemble
\begin{equation}
[\frac{\delta}{\delta N^{i}_{x\text{T}}},\frac{\delta}{\delta N^{j}_{x^{\prime}\text{T}}}]=[\frac{\delta}{\delta N^{i}_{x}},\frac{\delta}{\delta N^{j}_{x^{\prime}}}]+[\frac{\delta}{\delta \tilde{N}^{i}_{x}},\frac{\delta}{\delta \tilde{N}^{j}_{x^{\prime}}}].
\end{equation}
Since, by definition, the surface variables satisfy this algebra, we are left with
\begin{equation}
[\frac{\delta}{\delta \tilde{N}^{i}_{x}},\frac{\delta}{\delta \tilde{N}^{j}_{x^{\prime}}}]=-\frac{\delta}{\delta \tilde{N}^{i}_{x^{\prime}}}\partial_{jx}\delta(x,x^{\prime})-\frac{\delta}{\delta \tilde{N}^{j}_{x}}\partial_{ix}\delta(x,x^{\prime}),
\end{equation}
which is automatically satisfied for the generators (\ref{Ensemble Tangential Generators}) when the derivatives are given by (\ref{Ensemble Tangential Derivatives}).
\paragraph*{Tangential-Normal Lie bracket}
The Lie bracket that contains the normal and tangential generators characterizes the response of a normal deformation to diffeomorphisms of the surface. As this might suggest, knowing how something changes along a given surface depends entirely on the geometrical, or tensor, character of the object. Since the normal generator $\delta/\delta N_{x\text{T}}$ is a scalar density, any specific form of this generator that has this tensor character will satisfy the Lie bracket (\ref{Total Deformations tan-perp}).

To see how this Lie bracket splits into surface and ensemble pieces, we decompose the total generators accordingly. This time, however, the normal e-generator does depend on the surface variables (the LTFP equations contain the determinant of the metric $g^{1/2}$ and $\delta\Phi/\delta\tilde{N}_{x}=G_{x}[X^{\mu}]$ is assumed \emph{a priori} to have a surface variable dependence.), and thus there is not a complete separation of the ensemble and surface algebras as we had in the purely tangential case,
\begin{equation}
[\frac{\delta}{\delta \tilde{N}^{i}_{x}}+\frac{\delta}{\delta N^{i}_{x}},\frac{\delta}{\delta \tilde{N}_{x^{\prime}}}]=-\frac{\delta}{\delta \tilde{N}_{x}}\partial_{ix}\delta(x,x^{\prime}).
\label{Ensemble Tan-Perp LB}
\end{equation}
Heuristically we understand this because to correctly perform a Lie derivative of the normal generator $\delta/\delta\tilde{N}_{x}$ along the surface we must drag both the ensemble pieces $\rho$ and $\Phi$, as well as the $X^{\mu}$'s across the surface, and hence both tangential generators must be included.
\subsection*{Satisfying the dynamical Lie brackets}
Let us set the groundwork for satisfying the Lie bracket (\ref{Total Deformations perp-perp}). Again, we decompose this Lie bracket into surface and ensemble pieces, and as before, since the normal generators contain the surface variables, there is not a clean separation between the two algebras, as was the case in the purely tangential Lie bracket (\ref{Total Deformations tan-tan}). In fact, the exact relation becomes
\begin{align}
[\frac{\delta}{\delta \tilde{N}_{x}},\frac{\delta}{\delta \tilde{N}_{x^{\prime}}}]&+\frac{\delta}{\delta N_{x}}\frac{\delta}{\delta \tilde{N}_{x}}-\frac{\delta}{\delta N_{x^{\prime}}}\frac{\delta}{\delta \tilde{N}_{x}}\notag\\
&=-\left(g^{ij}_{x}\frac{\delta}{\delta \tilde{N}^{j}_{x}}+g^{ij}_{x^{\prime}}\frac{\delta}{\delta \tilde{N}^{j}_{x^{\prime}}}\right)\partial_{ix}\delta(x,x^{\prime})
\label{Total perp-perp mixed}
\end{align}
While this expression is imposing, we can employ a trick to make this more tractable. As was explicit in deriving these relations originally, the relations must hold, even if we act this operator upon an arbitrary functional $F_{\sigma}$. In particular, it must still hold for the functional $F_{\sigma}=\rho_{\sigma}$ in (\ref{Path Independence F}).\footnote{This notion of a necessary condition is consistent with the manner in which Teitelboim crafted his argument as well. Indeed, in the canonical framework, if we take the variables $\varphi_{nx}$ to be the canonical variables, then the necessary condition upon which Teitelboim's argument rests is \[\{\varphi_{ny},\{H_{\perp x},H_{\perp x^{\prime}}\}\},\] where the $H_{Ax}$ are a set of Hamiltonian generators roughly analogous to our $\delta/\delta N_{x\text{T}}^{A}$. The details are given in ch.4 of \cite{Teitelboim thesis}.} This is particularly convenient since, because we know the LTFP equations explicitly, the pieces that remain to be satisfied are largely known.

To be concise, let us abbreviate the LTFP equations as
\begin{equation}
\frac{\delta\rho_{\sigma}[\chi]}{\delta\tilde{N}_{x}}=E_{x}(\rho,\Phi;X^{\mu}).
\end{equation}
Applying the operator relation (\ref{Total perp-perp mixed}) to $\rho[\chi]=\rho$ and using
\begin{equation}
\frac{\tilde{\delta}\rho[\chi]}{\tilde{\delta}\rho[\chi^{\prime}]}=\delta[\chi-\chi^{\prime}]
\end{equation}
we get the relation \footnote{The ``mixed" terms in (\ref{Total perp-perp mixed}) that had both surface and ensemble generators cancel each other in this case since $E_{x}$ depends on the surface variables locally only through $g_{x}^{1/2}$. The resulting expression contains a delta function, and thus it is symmetric in $x$ and $x^{\prime}$, while (\ref{Total perp-perp mixed}) is antisymmetric $x-x^{\prime}$; they cancel.}
\begin{align}
&\int D\chi^{\prime} \left(E_{x}^{\prime}\frac{\tilde{\delta} E_{x^{\prime}}}{\tilde{\delta}\rho^{\prime}}+ G_{x}^{\prime}\frac{\tilde{\delta} E_{x^{\prime}}}{\tilde{\delta}\Phi^{\prime}}-E_{x^{\prime}}^{\prime}\frac{\tilde{\delta} E_{x}}{\tilde{\delta}\rho^{\prime}}- G_{x^{\prime}}^{\prime}\frac{\tilde{\delta} E_{x}}{\tilde{\delta}\Phi^{\prime}}\right)\notag\\
&=- \left(\frac{\delta\rho}{\delta\chi_{x}}g^{ij}_{x}\partial_{ix}\chi_{x}+\frac{\delta\rho}{\delta\chi_{x^{\prime}}}g^{ij}_{x^{\prime}}\partial_{ix^{\prime}}\chi_{x^{\prime}}\right)\partial_{jx}\delta(x,x^{\prime}),
\label{Ensemble perp-perp Lie Bracket rho-Reduced}
\end{align}
where we have used the notation that a prime denotes dependence on the field $\chi^{\prime}$ such that, for example, $\rho[\chi^{\prime}]=\rho^{\prime}$.
\paragraph*{Simplifications}
To make progress in solving the equation (\ref{Ensemble perp-perp Lie Bracket rho-Reduced}) for $G_{x}[\rho,\Phi;X^{\mu},\chi]$ we must proceed by making some reasonable assumptions. The first is the assumption of localizable fields. That is, if $G_{x}$ depends on the fields $\chi_{x}$ explicitly (as opposed to through $\rho$ and $\Phi$) then they should appear as local functions of $\chi_{x}$. The assumption of spatial locality is fairly reasonable given that the we are pursuing a theory of local evolution. As it turns out, no such assumption will need to be made for the surface variables, but they too will be of the local variety, and thus $G_{x}=G_{x}[\rho,\Phi;X^{\mu},\chi)$.

Furthermore, the space of possible ensemble functionals $G_{x}[\rho,\Phi;X^{\mu},\chi)$ is extraordinarily large. We make progress by searching for solutions in the simplest subspace; that of functions of $\rho$ and $\Phi$, such that $G_{x}=G_{x}(\rho,\Phi;X^{\mu},\chi)$. This is still, to be sure, an extremely large space consisting of all derivatives and all possible combinations of $\rho$ and $\Phi$, but it does not include, for example, integrals of those quantities. And in any case, this restriction leads to solutions for $G_{x}$, and thus for the evolution of $\Phi$, that are harmonious with the LTFP equations \ref{Local Time FP equation}) for $\rho$, which are also local in $\mathcal{C}$.
\paragraph*{Strategy}
A particularly appealing feature of these considerations is that the undetermined $G_{x}(\rho,\Phi;X^{\mu},\chi)$ is just a regular function, not a functional. This makes it reasonable to pursue a strategy where we expand $G_{x}$ in a series. This method is not elegant, but it is sufficient for our purposes here; perhaps future work will illuminate a better path forward in this regard. Nonetheless, it serves our current purpose to just formally expand $G_{x}$ in a series containing all the different manners that the inputs could be combined.

More explicitly the strategy is as follows. We first give a convenient ansatz for $G_{x}$ that simplifies the algebra, this ansatz is
\begin{equation}
-G_{x}=\frac{1}{2 \, g_{x}^{1/2}}\left(\frac{\delta\Phi}{\delta\chi_{x}}\right)^{2}+\frac{g_{x}^{1/2}}{2}g^{ij}_{x}\partial_{ix}\chi_{x}\,\partial_{jx}\chi_{x}+\frac{1}{g_{x}^{1/2}}\, M_{x}.
\label{Ansatz}
\end{equation}
We then insert this ansatz into the expression (\ref{Ensemble perp-perp Lie Bracket rho-Reduced}) and make use of the LTFP equations (\ref{Local Time FP equation}). The ansatz was specifically chosen so that many of the pieces would cancel in (\ref{Ensemble perp-perp Lie Bracket rho-Reduced}). The result is a condition only on the unknown $M_{x}$. From here, we perform the series expansion mentioned earlier. What allows us to make progress is that this equation should hold for \emph{arbitrary} choices of the inputs $\rho$, $\Phi$, etc. This means that terms that have the inputs appearing differently can be treated separately.\footnote{For example, as will become more apparent, we can inspect terms like $\delta^{4}\rho/\delta\chi_{x}^{2}\delta\chi_{x^{\prime}}^{2}$ separately from $(\delta^{2}\rho/\delta\chi_{x}\delta\chi_{x^{\prime}})^{2}$. This is a great simplification.} The result, as we shall see, is that a covariant ED is very restricted in the allowed dynamical models.
\paragraph*{Setup}
To proceed we must compute, in the expression (\ref{Ensemble perp-perp Lie Bracket rho-Reduced}), the four derivatives
\begin{align}
\frac{\tilde{\delta} E_{x^{\prime}}}{\tilde{\delta}\rho^{\prime}},\quad\frac{\tilde{\delta} E_{x^{\prime}}}{\tilde{\delta}\Phi^{\prime}},\quad\frac{\tilde{\delta} E_{x}}{\tilde{\delta}\rho^{\prime}},\quad\text{and}\quad\frac{\tilde{\delta} E_{x}}{\tilde{\delta}\Phi^{\prime}}.
\end{align}
Each of these are derivatives with respect to the variables $\rho$ and $\Phi$, which we call ensemble derivatives, or e-derivatives. Luckily, only two of these e-derivatives need be computed since the others are obtained by exchange of $x$ and $x^{\prime}$. Computing these e-derivatives when $E_{x}$ is given by the LTFP eqns.(\ref{Local Time FP equation}) gives \footnote{E-derivatives are a souped up version of standard partial derivatives from ordinary calculus. The construction we use here to compute these derivatives is the natural extension of functional derivatives to ensemble spaces. A good overview of standard functional derivatives is given in \cite{Greiner Field Quant}, from which the current procedure was inspired.}
\begin{equation}
\frac{\tilde{\delta} E_{x}}{\tilde{\delta} \rho^{\prime} }=-\frac{1}{g^{1/2}_{x}}\frac{\delta}{\delta\chi_{x}}\left(\delta\left[\chi-\chi^{\prime}\right]\frac{\delta\Phi}{\delta\chi_{x}}\right).
\label{Ensemble derivative E-rho'}
\end{equation}
and
\begin{equation}
\frac{\tilde{\delta} E_{x}}{\tilde{\delta} \Phi^{\prime}}=-\frac{1}{g^{1/2}_{x}}\frac{\delta}{\delta\chi_{x}}\left(\rho\frac{\delta}{\delta\chi_{x}}\delta\left[\chi-\chi^{\prime}\right]\right),
\label{Ensemble derivative E-Phi'}
\end{equation}
where the objects $\delta[\chi-\chi^{\prime}]$ are Dirac delta \emph{functionals}, which are the natural generalization of Dirac delta functions to the infinite dimensional configuration space $\mathcal{C}$.

As outlined above, we plug the derivatives (\ref{Ensemble derivative E-rho'}) and (\ref{Ensemble derivative E-Phi'}), as well as the ansatz (\ref{Ansatz}), into the condition (\ref{Ensemble perp-perp Lie Bracket rho-Reduced}). The Dirac delta functionals in (\ref{Ensemble derivative E-rho'}) and (\ref{Ensemble derivative E-Phi'}) collapse the integral in (\ref{Ensemble perp-perp Lie Bracket rho-Reduced}). The condition that results from these steps depends only on the undetermined $M_{x}$,
\begin{equation}
\frac{\delta}{\delta\chi_{x}}\left(\rho \, \frac{\delta}{\delta\chi_{x}} \, M_{x^{\prime}}\right)=\frac{\delta}{\delta\chi_{x^{\prime}}}\left(\rho \, \frac{\delta}{\delta\chi_{x^{\prime}}} \, M_{x}\right),
\label{Ensemble Constraint Equation}
\end{equation}
which is the relation we wish to work with.
\paragraph*{Solution}
We now look to expand $M_{x}$ in a series. Since at this point the surface variables $X^{\mu}$ do not enter into the calculations, we suppress the dependence of $M_{x}$ on $X^{\mu}$. A reasonable way to express the series is given by
\begin{equation}
M_{x}(\rho,\Phi;\chi)=\sum_{klmn} \, V^{(klmn)}_{x}(\chi_{x}) \, U^{(klmn)}(\rho,\Phi) \,\, \left(\frac{\delta^{(m)}\rho}{\delta\chi_{x}^{m}}\right)^{k} \,\, \left(\frac{\delta^{(n)}\Phi}{\delta\chi_{x}^{n}}\right)^{l},
\label{Ensemble Series Expansion}
\end{equation}
where $V_{x}^{(klmn)}(\chi_{x})$ is a regular distinct function of the fields $\chi_{x}$ for each $k$, $l$, $m$, and $n$, and the object $U^{(klmn)}(\rho,\Phi)$ is a regular function of the $\rho$'s and $\Phi$'s in a similar fashion. Inserting (\ref{Ensemble Series Expansion}) into (\ref{Ensemble Constraint Equation}), noting the Dirac delta identity
\begin{equation}
f(x) \, \delta(x,x^{\prime})=f(x^{\prime})\,\delta(x,x^{\prime}),
\end{equation}
and using the fact that (\ref{Ensemble Constraint Equation}) must hold for \emph{arbitrary} choices of $\rho$ and $\Phi$, one can show that $M_{x}$ must actually be of the form
\begin{equation}
M_{x}=V_{x}+U_{x}.
\end{equation}
Here we have that $V_{x}=V_{x}(\chi_{x})$ is a regular function of the $\chi_{x}$'s, while $U_{x}$ must be of the form
\begin{align}
U_{x}&=f_{1} \, \frac{\delta^{2}\rho}{\delta\chi_{x}^{2}}+f_{2} \, \frac{\delta^{2}\Phi}{\delta\chi_{x}^{2}}+\sum_{n=0}^{2} \, h_{n} \, \left(\frac{\delta\rho}{\delta\chi_{x}}\right)^{n}\left(\frac{\delta\Phi}{\delta\chi_{x}}\right)^{n-2},
\label{Definition U_x}
\end{align}
where each $f_{i}=f_{i}(\rho,\Phi)$ and $h_{n}=h_{n}(\rho,\Phi)$. A simple way to see the structure of (\ref{Definition U_x}) is to recall that in (\ref{Ensemble Constraint Equation}), that each side of the equation has two spatial indices $x$ and $x^{\prime}$, respectively. This implies that for each side in (\ref{Ensemble Constraint Equation}) to match the appropriate numbers of $x$'s and $x^{\prime}$'s, then $U_{x}$ should have exactly two spatial indices, as is the case in (\ref{Definition U_x}).

Plugging the $U_{x}$ in (\ref{Definition U_x}) into (\ref{Ensemble Constraint Equation}) can be used to find the undetermined functions $f_{i}(\rho,\Phi)$ and $h_{n}(\rho,\Phi)$. Indeed, one can show, after some laborious work, that we must have
\begin{equation}
f_{2}(\rho,\Phi)=h_{0}(\rho,\Phi)=h_{1}(\rho,\Phi)=0.
\end{equation}
The remaining functions $f_{1}=f_{1}(\rho)$ and $h_{2}=h_{2}(\rho)$ can be shown to depend only on $\rho$ and, in addition, are constrained to satisfy
\begin{equation}
\rho\frac{df_{1}(\rho)}{d\rho}+f_{1}(\rho)=0,
\end{equation}
and
\begin{equation}
h_{2}(\rho)=\frac{1}{2}\frac{df_{1}(\rho)}{d\rho}.
\end{equation}
The first equation is solved by
\begin{equation}
f_{1}(\rho)=\frac{1}{\rho}.
\end{equation}
Consequently $h_{2}(\rho)$ must be given by
\begin{equation}
h_{2}(\rho)=\frac{1}{2}\frac{df_{1}(\rho)}{d\rho}=-\frac{1}{2}\frac{1}{\rho^{2}}.
\end{equation}
Thus the solution we seek is given by
\begin{equation}
U_{x}=\lambda\left(\frac{1}{\rho}\frac{\delta^{2}\rho}{\delta\chi_{x}^{2}}-\frac{1}{2}\left(\frac{1}{\rho}\frac{\delta\rho}{\delta\chi_{x}}\right)^{2}\right),
\end{equation}
where the overall factor $\lambda$ fixes the units. This function is the so-called ``quantum potential," its importance will be discussed as we proceed to discuss the transition to quantum dynamics.
\paragraph*{Some discussion}
Putting everything together, the complete solution for the local normal evolution of $\Phi$ is given by\footnote{For notational simplicity we have re-defined \[\frac{V_{x}}{g_{x}^{1/2}}=g_{x}^{1/2}V^{\prime}_{x}\rightarrow g_{x}^{1/2} \, V_{x}.\]}
\begin{align}
-\frac{\delta\Phi}{\delta\tilde{N}_{x}}=\frac{1}{2 \, g_{x}^{1/2}}\left(\frac{\delta\Phi}{\delta\chi_{x}}\right)^{2}&+\frac{g_{x}^{1/2}}{2}g^{ij}_{x}\partial_{ix}\chi_{x}\,\partial_{jx}\chi_{x}+g_{x}^{1/2} \, V_{x}(\chi_{x})\notag\\
&+\frac{\lambda}{g_{x}^{1/2}} \, \left(\frac{1}{\rho}\frac{\delta^{2}\rho}{\delta\chi_{x}^{2}}-\frac{1}{2}\left(\frac{1}{\rho}\frac{\delta_{x}\rho}{\delta\chi_{x}}\right)^{2}\right),
\label{Local time HJ}
\end{align}
which we recognize as a set of local-time equations of the Hamilton-Jacobi type, in the context of scalar fields on curved spaces. For brevity, we will refer to these local equations as the ``Local time Hamilton-Jacobi" (LTHJ) equations.

Before moving on, an important point to notice is that, with the LTFP and LTHJ equations now known, we can speak unambiguously about the properties of the deformation algebra of the total generators. More directly, recall that the purely tangential Lie brackets separate cleanly between the ensemble and surface pieces. Perhaps surprisingly, something similar happens with the purely normal Lie brackets so that
\begin{equation}
[\frac{\delta}{\delta N_{x\text{T}}},\frac{\delta}{\delta N_{x^{\prime}\text{T}}}]=[\frac{\delta}{\delta N_{x}},\frac{\delta}{\delta N_{x^{\prime}}}]+[\frac{\delta}{\delta \tilde{N}_{x}},\frac{\delta}{\delta \tilde{N}_{x^{\prime}}}].
\end{equation}
Thus the only thing that keeps the total algebra from separating into a product of constituent surface and ensemble algebras is that the tangential-normal Lie bracket (\ref{Total Deformations tan-perp}) does not also behave in the same way. To be sure, this is not a problem in and of itself. However, in the next section, as we formulate the dynamics of $\rho$ and $\Phi$ as a Hamiltonian dynamics, the non-uniform way in which the surface and ensemble variables are handled will, so to speak, stick out like a sore thumb, and so will require ``fixing."
\section{Hamiltonian entropic dynamics}
\label{Section 6}
An interesting feature of the local-time evolution equations given by the LTFP eqns.(\ref{Local Time FP equation}) and the LTHJ eqns.(\ref{Local time HJ}) is that they lead naturally to a covariant Hamiltonian dynamics, and thus the e-phase space $\Gamma$ can indeed be regarded as a true phase space, being endowed with a conserved symplectic form, Poisson brackets, canonical transformations, etc. 

Our interest here in the connection to Hamiltonian dynamics from ED is many-fold. First, the derivation of a Hamiltonian structure from the seemingly basic principle of path independence is itself something novel and worth fleshing out for its own sake. From another point of view, the program of path independence developed by Teitelboim and Kucha\v{r} dealt almost exclusively with the canonical formalism, whereas our program has till now been silent in this regard. Thus making a connection between the two approaches seems apt. In fact, in pursuing this connection, there appear to be some discrepancies between the two approaches; we will find, however, these differences to be mostly \emph{superficial}.

One final reason for exploring this connection between ED and Hamiltonian dynamics is that it begins to lay the ground for understanding the formal relations between ED and quantum theory; the symplectic geometry of the Hamiltonian formalism being one of the necessary ingredients for recovering the natural K\"{a}hler geometry of quantum theory. In this section we develop these ideas a little further.
\subsection*{Hamiltonian generators}
To make the connection to Hamiltonian dynamics it is simplest to first rewrite the LTFP eqs.(\ref{Local Time FP equation}) in an alternative and very suggestive way involving the notion of an ensemble functional or \emph{e-functional}. Just as a regular functional such as $\rho \left[ \chi \right] $ maps a field $\chi $ into a real number (a probability in this case), an e-functional maps a functional, such as $\rho \left[ \chi \right] $ or $\Phi \left[ \chi \right] $, into a real number. Thus e-functionals can be thought of as ``functions" over the space $\Gamma$.\footnote{An excellent brief review of the ensemble calculus is given in the appendix of \cite{Hall/Reginatto 2003}.}
\paragraph*{Normal generator}
To proceed, introduce the e-functional $\tilde{H}_{\perp x}%
\left[ \rho,\Phi\right] $ such that, 
\begin{equation}
\frac{\delta \rho\left[ \chi \right] }{\delta \tilde{N}_{x}}=%
\frac{\tilde{\delta}\tilde{H}_{\perp x}\left[ \rho,\Phi
\right] }{\tilde{\delta}\Phi\left[ \chi \right] }
\label{Local Time FP equation H}
\end{equation}%
reproduces (\ref{Local Time FP equation}). We stress that writing (\ref{Local Time FP equation H}) does not involve any new assumptions; an appropriate $\tilde{H}_{\perp x}$ can always be found. Indeed, one can check that the appropriate $\tilde{H}_{\perp x}$ that reproduces (\ref{Local Time FP equation}) through (\ref{Local Time FP equation H}) is

\begin{equation}
\tilde{H}_{\perp x}\left[ \rho ,\Phi \right] =\int D\chi \,\rho \frac{1}{%
2g_{x}^{1/2}}\left( \frac{\delta \Phi }{\delta \chi _{x}}\right) ^{2}+\tilde{F}_{x}%
\left[ \rho \right] \,,  \label{Ensemble Hamiltonian F}
\end{equation}%
where $F_{x}[\rho ]$ is an \emph{a priori} undetermined integration constant (Note that, just as the normal e-generator contained the surface variables through the metric $g_{ij}(x)$, so too does the e-Hamiltonian $\tilde{H}_{\perp x}$ exhibit this dependence.).  However, for $\tilde{H}_{\perp x}$ to generate the correct dynamics for $\Phi$ as well as $\rho$, we must also choose
\begin{equation}
\tilde{F}_{x}=\int D\chi \, \rho \left(\frac{g_{x}^{1/2}}{2}g^{ij}\partial _{i}\chi _{x}\partial _{j}\chi _{x}+g_{x}^{1/2}V_{x}(\chi_{x})+\frac{4 \, \lambda}{g_{x}^{1/2}}\left(\frac{\delta\rho^{1/2}}{\delta\chi_{x}}\right)^{2}\right).
\end{equation}
The LTHJ equations can then be shown to be given by
\begin{equation}
\frac{\delta\Phi[\chi]}{\delta\tilde{N}_{x}}=-\frac{\tilde{\delta}\tilde{H}_{\perp x}[\rho,\Phi]}{\tilde{\delta}\rho[\chi]},
\label{Local Time HJ equation H}
\end{equation}
so that (\ref{Local Time FP equation H}) and (\ref{Local Time HJ equation H}) together form a set of functional covariant Hamilton's equations where $\rho$ plays the role of the canonical variable and $\Phi$ its conjugate momentum. We stress, however: this Hamiltonian system was not derived from any action principle, the dynamical equations come only from local entropic updating and path independence. This, we believe, requires a moment of appreciation. We have a developed a framework that seems to provide the necessary ingredients for when an action principle can be employed. And, although the context of scalar fields is not at all general, there is hope that a similar model can be used for more general situations.

Pressing on, however, as a result of these interesting developments, the normal e-generator from (\ref{Ensemble Normal Generators}) can be written as
\begin{equation}
\frac{\delta}{\delta\tilde{N}_{x}}=\int D\chi \left(\frac{\tilde{\delta}\tilde{H}_{\perp x}}{\tilde{\delta}\Phi\left[ \chi \right] }\frac{\tilde{\delta}}{\tilde{\delta}\rho[\chi]}-\frac{\tilde{\delta}\tilde{H}_{\perp x}}{\tilde{\delta}\rho\left[ \chi \right] }\frac{\tilde{\delta}}{\tilde{\delta}\Phi[\chi]}\right).
\label{Ensemble Hamiltonian Vector Field}
\end{equation}
This we identify as a Hamiltonian vector field that generates a spatially local flow normal to a given surface. Thus we can also append $\tilde{H}_{\perp x}$ with the interpretation of generating a spatially local evolution. And, as is typical of Hamiltonian vector fields, this flow preserves a symplectic form on the e-phase space $\Gamma$.
\paragraph*{Tangential Generators}
For a covariant theory done properly, it is not sufficient to speak simply of normal evolution, but also of the tangential deformations. Thus it must be possible to construct Hamiltonian generators that generate the appropriate tangential deformations. Indeed such a construction is possible. Consider the e-functional
\begin{equation}
\tilde{H}_{ix}[\rho,\Phi]=\int D\chi \, \Phi \frac{\delta\rho}{\delta\chi_{x}}\partial_{ix}\chi_{x},
\label{Ensemble Momentum}
\end{equation}
which we will call the ensemble momentum or e-momentum. It is not difficult to show that taking an e-derivative gives
\begin{equation}
\frac{\delta\rho[\chi]}{\delta \tilde{N}_{x}^{i}}=\frac{\tilde{\delta}\tilde{H}_{ix}[\rho,\Phi]}{\tilde{\delta}\Phi[\chi]}=\frac{\delta\rho}{\delta\chi_{x}}\partial_{ix}\chi_{x}.
\end{equation}
This gives a half of the tangential version of Hamilton's equations. The other half is given by performing an e-derivative with respect to $\rho$, and so we have\footnote{To compute this derivative we must perform an integration by parts and use the Dirac delta identity\[\frac{\delta}{\delta\chi_{x}}\;\partial_{ix}\chi_{x}=0.\]}
\begin{equation}
\frac{\delta\Phi[\chi]}{\delta \tilde{N}_{x}^{i}}=-\frac{\tilde{\delta}\tilde{H}_{ix}[\rho,\Phi]}{\tilde{\delta}\rho[\chi]}=\frac{\delta\Phi}{\delta\chi_{x}}\partial_{ix}\chi_{x}.
\end{equation}
This means that the tangential e-generator can be written using the e-momentum to give 
\begin{equation}
\frac{\delta}{\delta\tilde{N}_{x}^{i}}=\int D\chi \left(\frac{\tilde{\delta}\tilde{H}_{i x}}{\tilde{\delta}\Phi\left[ \chi \right] }\frac{\tilde{\delta}}{\tilde{\delta}\rho[\chi]}-\frac{\tilde{\delta}\tilde{H}_{i x}}{\tilde{\delta}\rho\left[ \chi \right] }\frac{\tilde{\delta}}{\tilde{\delta}\Phi[\chi]}\right).
\label{Ensemble Momentum Vector Field}
\end{equation}
Thus the set of Hamiltonian e-functionals $\tilde{H}_{Ax}=(\tilde{H}_{\perp x},\tilde{H}_{ix})$ form a set local generators analogous in function to the operators $\delta/\delta\tilde{N}_{x}^{A}$.
\paragraph*{The Poisson bracket}
In the standard language, functions (or here e-functionals) over the relevant phase space are often called ``observables." For each such observable $\tilde{Y}[\rho,\Phi]$ and $\tilde{Z}[\rho,\Phi]$ over the e-phase space $\Gamma$ there corresponds a Hamiltonian vector field denoted $\bar{v}_{\tilde{Y}}$ and $\bar{v}_{\tilde{Z}}$, respectively, which are analogous in form to (\ref{Ensemble Hamiltonian Vector Field}) and (\ref{Ensemble Momentum Vector Field}). It is well known\footnote{See \cite{Schutz} for a good introduction to geometrical methods in physics. Ch. 5 has a nice overview of the structure of Hamiltonian vector fields.} that the Lie brackets of these Hamiltonian vector fields themselves form a Lie algebra given by
\begin{equation}
[\bar{v}_{\tilde{Y}},\bar{v}_{\tilde{Z}}]=\bar{v}_{\{\tilde{Y},\tilde{Z}\}},
\end{equation}
where the vector $\bar{v}_{\{\tilde{Y},\tilde{Z}\}}$ is the vector field that arises from the the bilinear product $\{\tilde{Y},\tilde{Z}\}$ of $\tilde{Y}$ and $\tilde{Z}$, called the Poisson bracket. This bracket is given by
\begin{equation}
\{\tilde{Y},\tilde{Z}\}=\int D\chi\left(\frac{\tilde{\delta}\tilde{Y}}{\tilde{\delta}\rho[\chi]}\frac{\tilde{\delta}\tilde{Z}}{\tilde{\delta}\Phi[\chi]}-\frac{\tilde{\delta}\tilde{Y}}{\tilde{\delta}\Phi[\chi]}\frac{\tilde{\delta}\tilde{Z}}{\tilde{\delta}\rho[\chi]}\right),
\end{equation}
which itself defines an algebra between the observables $\tilde{Y}$ and $\tilde{Z}$ on $\Gamma$ that is analogous to the Lie brackets between $\bar{v}_{\tilde{Y}}$ and $\bar{v}_{\tilde{Z}}$.\footnote{In fact, the relationship between the two is a homomorphism. See \cite{Isham/Kuchar I} for a more detailed discussion.}

The Poisson bracket operation satisfies the properties, such as the Leibniz rule, that are characteristic of differentiations. In fact, one can think of the Poisson bracket $\{\tilde{Y},\tilde{Z}\}$ as being \textquotedblleft the derivative of $\tilde{Y}$ with respect to the flow generated by $\tilde{Z}$.\textquotedblright  \ Thus this allows us to write our Hamilton's equations in the succinct form
\begin{equation}
\left\{\rho[\chi],\tilde{H}_{Ax}\right\}=\frac{\tilde{\delta}\tilde{H}_{Ax}}{\tilde{\delta}\Phi[\chi]}=\frac{\delta\rho[\chi]}{\delta\tilde{N}^{A}_{x}}\quad\text{and}\quad\left\{\Phi[\chi],\tilde{H}_{Ax}\right\}=-\frac{\tilde{\delta}\tilde{H}_{Ax}}{\tilde{\delta}\rho[\chi]}=\frac{\delta\Phi[\chi]}{\delta\tilde{N}^{A}_{x}}.
\end{equation}
More generally, for a generic e-functional $\tilde{Y}[\rho,\Phi]$ defined only on $\Gamma$ (i.e. no surface variables, this case will be discussed later) its response to a deformation $\delta\xi^{A}_{x}$ can be given using the language of Poisson brackets by
\begin{equation}
\delta \tilde{Y}=\int dx \left\{\tilde{Y},\tilde{H}_{Ax}\right\} \, \delta\xi^{A}_{x}.
\label{Ensemble variation PB}
\end{equation}
If the e-functional $\tilde{Y}=\tilde{Y}[\rho,\Phi;X^{\mu}]$ were to also include a geometry dependence, then the total variation must be modified as
\begin{equation}
\delta \tilde{Y}=\int dx \left(\left\{\tilde{Y},\tilde{H}_{Ax}\right\}+\frac{\delta\tilde{Y}}{\delta N^{A}_{x}}\right) \, \delta\xi^{A}_{x},
\label{Ensemble variation PB+Surface}
\end{equation}
to account for this addition.
\subsection*{Canonical path independence}
Having constructed a set of e-Hamiltonian generators and a notion of an algebra between them, we now have the formal tools to discuss path independence in the canonical setting. The existence of a mapping between the Lie bracket and Poisson bracket algebras suggests that we might be able to \emph{represent} the Lie brackets between the e-generators $\delta/\delta\tilde{N}^{A}_{x}$ as a set of Poisson brackets between the corresponding e-Hamiltonian generators. Indeed, if we take
\begin{equation}
\bar{v}_{\tilde{H}_{Ax}}\equiv \bar{v}_{Ax}=\frac{\delta}{\delta\tilde{N}^{A}_{x}}
\end{equation}
as the Hamiltonian flows generated by the observables $\tilde{H}_{Ax}$, then the Lie brackets among the $\bar{v}_{Ax}$ in turn correspond to a set of Poisson brackets among the $\tilde{H}_{Ax}$'s themselves. This set of Poisson brackets is given by
\begin{subequations}
\begin{eqnarray}
&&\left\{ \tilde{H}_{\bot x},\tilde{H}_{\bot x^{\prime }}\right\} =\left(
g_{x}^{ij}\tilde{H}_{ix}+g_{x^{\prime }}^{ij}\tilde{H}_{ix^{\prime }}\right) \partial
_{jx}\delta (x,x^{\prime })  \label{e-PB 1} \\
&&\left\{ \tilde{H}_{ix},\tilde{H}_{\bot x^{\prime }}\right\}-\frac{\delta\tilde{H}_{\perp x^{\prime}}}{\delta N^{i}_{x}} =\tilde{H}_{\bot x}\partial _{ix}\delta(
x,x^{\prime })  \label{e-PB 2} \\
&&\left\{ \tilde{H}_{ix},\tilde{H}_{jx^{\prime }}\right\}  =\tilde{H}_{ix^{\prime }}\,\partial
_{jx}\delta (x,x^{\prime })+\tilde{H}_{jx}\,\partial _{ix}\delta (x,x^{\prime }).
\label{e-PB 3}
\end{eqnarray}
\end{subequations}
The eqns.(\ref{e-PB 1}-\ref{e-PB 3}) require some comments. First, as was noted above, the total Lie bracket algebras for the surface and ensemble generators do not decompose into a product, and thus the situations for the Poisson brackets of the e-Hamiltonians also does not. Additionally, the equation (\ref{e-PB 2}), which is the relation that is responsible for the mixing of the surface and ensemble algebras, has a peculiar term with a derivative. This peculiar term is a direct result of the fact that $\tilde{H}_{\perp x}$ contains surface variables, and thus invokes the use of (\ref{Ensemble variation PB+Surface}); the minus sign results from the sign convention of the Poisson brackets.
\paragraph*{Dirac parameterized dynamics}
Although there is nothing, in principle, incorrect about the eqns.(\ref{e-PB 1})-(\ref{e-PB 3}), some objections \emph{can}, however, be raised about them. One objection is on aesthetic grounds. The appearance of the surface generators as occurring on a different footing from the e-Hamiltonians makes manipulations more cumbersome and the expressions less concise and elegant. Another criticism comes from the viewpoint of looking to incorporate a dynamical theory of gravity into this formalism. There, the geometry of the surface, encoded by the metric $g_{ij}(x)$, is itself dynamical. As is well known,\footnote{A good overview of this issue is given in the introduction of \cite{Kuchar 1972}.} the metric components $g_{ij}(x)$ contain true dynamical information, as well as kinematical content equivalent to that given by the surface variables. As a consequence, when one treats the metric field $g_{ij}$ as a canonical variable in the geometrodynamics approach to dynamical gravity, the kinematical variables too are also \emph{automatically} included as part of the relevant phase space.

These points suggests that a natural way to proceed --- that is both economical and more compliant with future generalization --- is to follow Dirac \cite{Dirac 1951}\cite{Dirac Lectures} in promoting the surface variables $X^{\mu}_{x}$, or equivalently $X^{A}_{x}$, themselves into canonical variables.\footnote{This ``trick" due to Dirac is analogous to allowing the time $t$ to become a canonical variable in particle mechanics.}\footnote{Since the surface variables $X^{\mu}_{x}$ simply label points in spacetime, it makes no difference whether we use the general spacetime index $\mu=(0,1,2,3)$ or the label $A=(\perp,1,2,3)$ to do this; however, the conjugate momenta $\pi _{A}(x)=\pi_{Ax}=(\pi_{\perp x},\pi_{ix})$ of the latter set of variables will be more useful for our purposes.} Such a maneuver allows us to fully represent eq.(\ref{e-PB 2}) by Poisson brackets and Hamiltonian generators. This is achieved by formally introducing the auxiliary variables $\pi _{A}(x)=\pi_{Ax}=(\pi_{\perp x},\pi_{ix})$ that will play the role of momenta conjugate to $X_{x}^{A}$. These $\pi $'s are defined strictly through the Poisson bracket relations, 
\begin{equation}
\{ X_{x}^{A },X_{x^{\prime }}^{B }\}=0\,,\quad \{ \pi_{A x},\pi _{B x^{\prime }}\}=0\,,\quad \{ X_{x}^{A },\pi
_{B x^{\prime }}\}=\delta _{B }^{A }\delta (x,x^{\prime }),
\label{PB HS}
\end{equation}
where the Poisson brackets are defined over the extended phase space $\Gamma^{*}$ with coordinates $(\rho,\Phi;X^{A},\pi_{A})$. This extended Poisson bracket is explicitly given as
\begin{align}
\left\{ \tilde{Y},\tilde{Z}\right\} =&\int D\chi \, \left( \frac{\tilde{\delta} \tilde{Y}}{\tilde{\delta} \rho }\frac{%
\tilde{\delta} \tilde{Z}}{\tilde{\delta} \Phi }-\frac{\tilde{\delta} \tilde{Y}}{\tilde{\delta} \Phi }\frac{\tilde{\delta} \tilde{Z}}{\tilde{\delta}\rho }\right)\notag\\
&+\int dx\,\left( \frac{\delta \tilde{Y}}{\delta X_{x}^{A }}\frac{%
\delta \tilde{Z}}{\delta \pi _{A x}}-\frac{\delta \tilde{Y}}{\delta \pi _{A x}}\frac{\delta \tilde{Z}}{\delta X_{x}^{A }}\right) ,
\label{General PB}
\end{align}%
where $\tilde{Y}$ and $\tilde{Z}$ are arbitrary functionals depending on the inputs $(\rho,\Phi;X^{A},\pi_{A})$.

To understand further the role of the $\pi$'s, consider a functional $\tilde{Y}$ depending on just $\rho$, $\Phi$, and the surface variables $X^{A}_{x}$. Compute now its Poisson bracket with the conjugate momentum $\pi_{Ax}$. Since $\tilde{Y}$ depends only on the $X^{A}_{x}$ (not the $\pi$'s and $X$'s), this gives
\begin{equation}
\{\tilde{Y},\pi_{Ax}\}=-\frac{\delta\tilde{Y}}{\delta X^{A}_{x}}\quad\text{which is the same as}\quad \{\tilde{Y},\pi_{Ax}\}=-\frac{\delta\tilde{Y}}{\delta N^{A}_{x}},
\end{equation}
using our established, familiar notation. Thus from this we conclude that the role of the conjugate momenta $\pi_{Ax}$ is that of a Hamiltonian generator that is analogous to the surface generators
\begin{equation}
\pi_{Ax}\leftrightarrow \frac{\delta}{\delta N^{A}_{x}}.
\end{equation}
\paragraph*{Path independence, algebra, and constraints}
Having extended the phase space to include the surface variables, it is now possible to write (\ref{e-PB 2}) in a more streamlined manner as
\begin{equation}
\left\{ \tilde{H}_{ix}+\pi_{ix},\tilde{H}_{\bot x^{\prime }}\right\} =\tilde{H}_{\bot x}\partial _{ix}\delta(x,x^{\prime }).
\label{e-PB 2 fix}
\end{equation}
Indeed, the form of (\ref{e-PB 2 fix}) suggests a further step. We introduce now the notion of a total Hamiltonian generator, given by
\begin{equation}
H_{Ax}=\pi_{Ax}+\tilde{H}_{Ax}.
\label{Total Hamiltonian Generator}
\end{equation}
It is now possible to rewrite the Lie bracket algebra (\ref{Total Deformations perp-perp})-(\ref{Total Deformations tan-tan}) of the total generators in a tidy way using a Poisson bracket algebra of the corresponding Hamiltonian generators
\begin{subequations}
\begin{eqnarray}
\left\{ H_{\perp x},H_{\perp x^{\prime}}\right\} &=&\left(
g_{x}^{ij}H_{ix}+g_{x^{\prime }}^{ij} \, H_{ix^{\prime }}\right) \partial
_{jx}\delta (x,x^{\prime })  \label{Total PB 1} \\
\left\{ H_{ix},H_{\bot x^{\prime }}\right\} &=&\tilde{H}_{\bot x}\partial _{ix}\delta(
x,x^{\prime })  \label{Total PB 2} \\
\left\{ H_{ix},H_{jx^{\prime }}\right\}  &=& H_{ix^{\prime }}\,\partial
_{jx}\delta (x,x^{\prime })+H_{jx}\,\partial _{ix}\delta (x,x^{\prime }),
\label{Total PB 3}
\end{eqnarray}
\end{subequations}
which agrees completely with the Poisson bracket algebra obtained by DKT.

While this step completes the formal ability to write the algebra of path independence in the canonical language, our work is not yet complete. The astute reader may have noticed that the total Hamiltonians $H_{Ax}$ are defined over a space $\Gamma^{*}$ that is \emph{larger} than the original space of interest $\mathcal{T}=\Sigma \, \otimes \, \Gamma $, owing to the inclusion of the momenta $\pi_{Ax}$. To remedy this issue we must supplement the algebra (\ref{Total PB 1})-(\ref{Total PB 3}) with an additional set of constraints
\begin{equation}
H_{Ax}\approx 0,
\label{Hamiltonian constraints}
\end{equation}
where `` $\approx$ " is a weak equality, in the sense of Dirac. If we had pursued path independence from the canonical perspective from the outset, as was done by Teitelboim \cite{Teitelboim 1972}\cite{Teitelboim thesis}, then our result would have been exactly the algebra (\ref{Total PB 1})-(\ref{Total PB 3}) with the additional constraints (\ref{Hamiltonian constraints}). And so, we have shown, in essence, the formal equivalence between our approach and that of Teitelboim and Kucha\v{r} (and also of Dirac).

It is important to note, however, that at this level of dynamics, with a fixed background geometry, the Hamiltonian constraints are more of a triviality; they are a result of formalism, rather than of genuine physical content. Thus the ability to work with or without them is not terribly important; the same dynamics is obtained for $\rho$ and $\Phi$ regardless of these constraints. Obviously when one moves to a dynamical geometry, the constraints analogous to (\ref{Hamiltonian constraints}) are of fundamental importance and we may have to revise our framework accordingly. Such matters are, however, a topic for future work.

Nonetheless, what we stress here is not the Hamiltonian constraints \emph{per s\'{e}}, but only that we have achieved results that are equivalent to Teitelboim and company, and with a minimum of extra kinematical structure (i.e. without the $\pi$'s) to boot. An additional benefit of our approach is that we avoid here the necessity to argue for a Hamiltonian formalism in the first place, unlike DKT,\footnote{The exact quote is this \cite{Teitelboim 1980}
\begin{quote}
\textit{We shall assume that such an evolution law may be given in Hamiltonian form. This is quite a strong assumption, but it looks like a reasonable one. Not least because the final product based on this assumptions looks so simple and natural that one can hardly doubt the cogency of the assumption.}
\end{quote}} or past efforts at a covariant ED \cite{Ipek/Abedi/Caticha 2017}. Our approach utilizes entropic updating, and past that, the formalism of flows generated by vector fields is rather general with much weaker assumptions.
\subsection*{Covariant dynamical models}
Thus far we have not discussed the particular dynamical models that are allowed by path independence. Although the LTFP equations are fixed by entropic updating, there is still some freedom left in choosing what ``potentials" we allow in the LTHJ equations.
\paragraph*{Hybrid models}
The simplest covariant theories we can obtain fall under what we call ``hybrid" models of ED. These are models of ED where the scalar fields still undergo a Brownian motion, but they are guided by a completely classical HJ functional $\Phi[\chi]$.\footnote{See \cite{Bartolomeo et al 2014} for a more detailed discussion of these Hybrid models, which may be relevant for studying the classical limit, or even as models for classical systems interacting with quantum systems, such as in \cite{Hall/Reginatto 2005}.} These hybrid models are obtained when $\Phi$ satisfies ($\lambda=0$)
\begin{equation}
-\frac{\delta \Phi }{\delta \tilde{N}_{x}}=\frac{1}{2 \, g_{x}^{1/2}}\left( \frac{\delta \Phi }{\delta \chi _{x}}\right)^{2}
+\frac{g_{x}^{1/2}}{2}g^{ij}\partial _{i}\chi _{x}\partial _{j}\chi _{x}+g_{x}^{1/2} \, V_{x}(\chi_{x}).
\label{Local time HJ Hybrid}
\end{equation}%
When we have $V_{x}=0$ we identify this as the LTHJ that one obtains in a Klein-Gordon model for free massless relativistic scalar field. Choosing for the potential \[V_{x}(\chi_{x})=\frac{1}{2}m^{2} \, \chi_{x}^{2}\] introduces a mass to the field and \[V_{x}(\chi_{x})=\sum_{n}\frac{c^{(n)}}{n!} \, \chi_{x}^{n}\] introduces ever more complicated self interactions into the picture. A lot of work remains to be done in exploring these Hybrid models, but they may prove to be interesting models for ``mesoscopic" phenomenon that are something between classical and quantum.\footnote{Unfortunately the word ``hybrid" to describe these theories can be slightly confusing as the word ``hybrid" is also used in the literature for theories describing the coupling of classical and quantum systems, called \textit{classical-quantum hybrids} (see \cite{Hall/Reginatto 2005} and references therein). We stress here that this model does not couple two systems -- one classical and the other quantum --- but rather, it is a single system that exhibits characteristics of both.}
\paragraph*{Quantum dynamics}
While hybrid models are interesting in their own right, they do not seem to be the ``correct" models for describing nature at a fundamental level. What is particularly interesting about our current approach is that the only other qualitatively distinct dynamical model that we could construct, is a \emph{quantum} mechanical one.

For our current purposes, it is more instructive to look at the e-Hamiltonian $\tilde{H}_{\perp x}$ rather than the LTHJ equations directly. For simplicity, let us consider the $\tilde{F}_{x}[\rho]$ where we have set the potential terms $V_{x}(\chi_{x})=0$, but allow $\lambda\neq 0$, unlike the hybrid models, so that we have
\begin{align}
\tilde{H}_{\perp x}\left[ \rho ,\Phi \right] =\int D\chi \,\rho  \left(\frac{1}{2g_{x}^{1/2}}  \left( \frac{\delta \Phi }{\delta \chi _{x}}\right) ^{2} +\frac{4 \, \lambda}{g_{x}^{1/2}} \left(\frac{\delta\rho^{1/2}}{\delta\chi_{x}}\right)^{2}+\frac{g_{x}^{1/2}}{2}g^{ij}\partial _{i}\chi _{x}\partial _{j}\chi _{x}\right).
\label{Ensemble Hamiltonian Free Quantum}
\end{align}%
Since the functional $\Phi[\chi]$ and the probability $\rho[\chi]$ both appear, essentially, quadratically, we may suppose that there is a transformation that simplifies the appearance of the Hamiltonian. Consider now the canonical transformation (also called a \emph{Madelung} transformation) to the variables
\begin{equation}
\Psi_{k}[\chi]=\rho^{1/2}[\chi]e^{i k \, \Phi[\chi]/\eta}\quad\text{and}\quad\Psi^{*}_{k}[\chi]=\rho^{1/2}[\chi]e^{-i k \, \Phi[\chi]/\eta}.
\label{Definition Psi}
\end{equation}
This transformation defines a one parameter group of canonical transformations labeled by the parameter $k$. For an arbitrary choice of $k$ the coupled dynamics of the LTFP and LTHJ equations will be non-linear and consequently difficult to solve. But, just as one performs a canonical transformation in Hamilton-Jacobi theory to make the dynamical problem as simple as possible, a similar tactic can be employed here. That is to say, we look for a choice of $k$ that makes the dynamics as simple as possible --- we choose $k$ so that the Hamiltonian $\tilde{H}_{\perp x}$ becomes \emph{linear}. Indeed, such an optimal choice is possible and is given by
\begin{equation}
\hat{k}=\left(\frac{\eta^{2}}{8\lambda}\right)^{1/2}.
\end{equation}
This suggests defining an optimal scale to our theory, given by $\eta/\hat{k}$, that can be associated with Planck's constant $\hbar$
\begin{equation}
\frac{\eta}{\hat{k}}=\left(8\,\lambda\right)^{1/2}=\hbar,
\end{equation}
which puts the e-Hamiltonian (\ref{Ensemble Hamiltonian Free Quantum}) in the tidy form
\begin{align}
\tilde{H}_{\perp x}\left[ \rho ,\Phi \right] =\int D\chi \left(\frac{\hbar^{2}}{2\, g_{x}^{1/2}}\frac{\delta\Psi^{*}[\chi]}{\delta\chi_{x}}\frac{\delta\Psi[\chi]}{\delta\chi_{x}}+ \frac{g_{x}^{1/2}}{2}g^{ij}\partial _{i}\chi _{x}\partial _{j}\chi _{x}\Psi^{*}[\chi]\Psi[\chi]\right).
\end{align}
Thus the transformed variables $\Psi_{\hat{k}}[\chi]=\Psi[\chi]$ with optimal parameter $\hat{k}$ satisfy a \emph{single} set of linear dynamical equations (for $V_{x}\neq 0$)
\begin{equation}
i\hbar \frac{\delta \Psi }{%
\delta \tilde{N}_{x}}=-\frac{1}{g_{x}^{1/2}}\frac{\hbar ^{2}}{2}\frac{\delta ^{2}\Psi }{\delta \chi _{x}^{2}}+\frac{g_{x}^{1/2}}{2}g^{ij}\partial _{i}\chi_{x}\,\partial _{j}\chi _{x}\,\Psi+g_{x}^{1/2}V_{x}\,\Psi,
\end{equation}
which we identify as a set of local time Schr\"{o}dinger equations. The remaining generators and quantities in the theory can all be expressed in these new variables, and so we make the claim that this is, in fact, a covariant quantum theory.
\paragraph*{Comments on quantum theory}
What we have shown here is that Schr\"{o}dinger time evolution arises quite naturally within the framework of covariant ED. In particular, the quantum potential is derived, making quantum theory one of the few possible modes of dynamics in this framework. This is important. There are many frameworks --- Nelson's stochastic mechanics \cite{Nelson}
\cite{Nelson Book}, Bohmian mechanics \cite{Bohm/Hiley 1987}\cite{Bohm/Hiley 1989}, the hydrodynamical formulation of quantum mechanics \cite{Madelung}\cite{Takabayasi}, etc. --- that bare some formal resemblance to ED in that the main dynamical objects are the real-valued quantities $\rho$ and $\Phi$, not the complex $\Psi$ and $\Psi^{*}$. Naturally, all of these frameworks arrive at the same dynamical equations, albeit using different methods and assigning different interpretations. However, for each such framework to recover quantum dynamics, they must inevitably include the quantum potential, and \emph{all} of these approaches make \emph{serious} assumptions regarding the quantum potential: either they argue only heuristically for it or assume it outright. Here the quantum potential is selected through a system of constraints. The result is rigorous, and seems to be one of the only such derivations of the quantum potential we have found in the literature.

On another note, the developments here show that quantum dynamics is a special case of Hamiltonian dynamics, a result that was known already due to the works of Kibble \cite{Kibble}, Heslot \cite{Heslot}, as well as Ashtekar and Fulling \cite{Ashtekar}. Yet none of these works \textit{derive} this relationship, it is simply posited by \emph{fiat}. And while, in many ways, the current work formally resembles that of Hall and Reginatto  \cite{Hall/Reginatto 2002}, even there an action principle was used to generate the Hamiltonian dynamics. Here, on the other hand, the Hamiltonian dynamical structure of quantum theory is \emph{derived} from first principles.

Still, quantum theory is actually much richer than just the symplectic structure inherent in the Hamiltonian framework. Quantum theory has a K\"{a}hler structure as well \cite{Cirelli I} --- that is, it has a complex structure and a Riemannian metric in addition to a symplectic form. Although the complex structure does also arise here somewhat naturally --- as seen by the efficacy of the transformation to the complex variables $\Psi$ and $\Psi^{*}$ --- the Riemannian geometry seems to be missing. This is, however, not quite the full picture. There is, in fact, a natural metric structure --- information geometry --- that comes associated with the probability $\rho$, half the e-phase space.\footnote{An accessible introduction to information geometry is given in ch. 7 of \cite{Caticha 2012}. A more in-depth discussion of these topics is given by Amari in \cite{Amari 1985}.} In \cite{Reginatto Kahler 2014} Reginatto utilizes this structure, together with some natural compatability conditions to extend the metric structure to the full phase space. This scheme then does indeed reproduce the full scope of quantum theory. It would seem a similar construction is possible here as well and is the subject of future works.\footnote{Indeed, in recent work by A. Caticha \cite{Caticha 2017}, he produces, albeit in a different context, a very natural argument for K\"{a}hler geometry in ED from the consideration of information geometry.}
\section{Conclusion and Discussion}
\label{Section 7}
Covariant ED provides an alternative method for deriving quantum theory from the basic principles of entropic inference and path independence. The goal, which we view as successful in the current case, is to discard completely the black box that is ``quantization" and replace it with a set of well defined constraints, given here by (\ref{Constraint 1}) and (\ref{Constraint 2}) together with (\ref{Total Deformations perp-perp})-(\ref{Total Deformations tan-tan}).

In stages, the work here has shown:
\begin{itemize}
\item The ability to construct a manifestly covariant ED.
\item That entropic updating and path independence alone serve to \emph{derive} a Hamiltonian dynamics.
\item That quantum theory stands, nearly uniquely, as the only way to formulate a covariant ED.
\end{itemize}

The method contributes from both technical and conceptual viewpoints. Using primarily the tools we associate with classical physics --- vector field flows, and later Hamiltonians --- we were able to model a theory that is intrinsicly statistical and quantum. This allows us to avoid annoyances such as operator ordering ambiguities that these conventional quantization methods come with. Indeed, many of the problems associated with the Dirac quantization method \cite{Dirac Lectures} and laborious techniques necessary to implement it (such as the identification and elimination of second-class constraints etc.) are completely sidestepped.

The principal benefits to our approach, however, are conceptual. The wave functional $\Psi$ has a well defined role as an epistemic quantity that expresses a state of knowledge. The field $\chi_{x}$, on the other hand, has a privleged ontological status; in the language of John Bell, they are the \emph{beables} of the theory. This privleged role, together with epistemic status of $\Psi$, allows for a solution of the quantum measurement problem \cite{Caticha/Johnson}\cite{Caticha Vanslette 2017} as an instance of discrete entropic updating through Bayes rule. This is especially relevant in the current context, that of quantum fields on curved spaces, since the normal notion of a particle fails (see  \cite{Fulling}\cite{Davies}\cite{Unruh}\cite{Birrell/Davies}), while the field concept remains. Thus covariant ED seems to have a toolkit that is naturally suited to solving the age old question of, what is a particle?

Finally, to recap, the work here has presented several dynamical models for inferring the behavior a scalar field $\chi_{x}$. Updating the probability $\rho[\chi]$ in response to a background spacetime geometry and drift potential $\phi[\chi]$ leads to a diffusion. Relaxing the assumption of a background drift potential $\phi[\chi]$, or equivalently $\Phi[\chi]$, in favor of a dynamical one leads to more robust dynamical models: quantum theory. This then begs the question, what happens when we allow the geometry itself to also become dynamical?
\section*{Acknowledgements}

The author would like to provide special thanks to M. Abedi and to my advisor A. Caticha, both of whom were instrumental in developing this work, and to the University at Albany and its physics department for supporting me and my research; their helpfulness and care cannot be overstated. Additional thanks go to D. Bartolomeo, N. Carrara, S. Nawaz, P. Pessoa, K. Vanslette, and A. Yousefi for many stimulating discussions on entropy, inference and quantum mechanics.

\appendix\numberwithin{equation}{section}

\section{The local-time Fokker-Planck equations}

\label{appendix FP}To rewrite the dynamical equation (\ref{Evolution
equation}) in differential form consider the probability $P[\chi ,\sigma|\chi _{0},\sigma _{0}]\equiv P[\chi|\chi_{0}]$ of a \emph{finite} transition from a field configuration $\chi _{0}$ at some early surface $\sigma _{0}$ to a configuration $\chi $ at a later $\sigma $; we employ the notation $\chi\to(\chi,\sigma)$ to represent both configurational and surface dependence. The result of a further
evolution from $\sigma $ to a neighboring $\sigma ^{\prime }$ obtained from $\sigma $ by an infinitesimal normal deformation $\delta \xi^{\perp}_{x}$ is given by (\ref{Evolution equation}), 
\begin{equation}
P[\chi^{\prime}|\chi_{0}] =\int D\chi \,P[\chi^{\prime}|\chi] P[\chi|\chi_{0}] ~.
\end{equation}
To obtain a differential equation one cannot just Taylor expand as $\delta\xi _{x}^{\bot }\rightarrow 0$ because $P[\chi^{\prime}|\chi]$ becomes a very singular object --- a delta functional. Instead, we multiply by an arbitrary smooth test functional $T\left[ \chi ^{\prime }\right] $ and integrate 
\begin{equation}
\int D\chi ^{\prime }\,P[\chi^{\prime}|\chi_{0}] T\left[ \chi ^{\prime }\right] =\int D\chi \,P[\chi|\chi_{0}] \int D\chi ^{\prime }\,T\left[\chi ^{\prime }\right] \,P[\chi^{\prime}|\chi] \,. 
\label{Test Function a}
\end{equation}%
Next expand the test function $T\left[ \chi ^{\prime }\right] $ about $\chi $ in powers of $\Delta \chi =\chi ^{\prime }-\chi $. Since we deal here with a Brownian motion, to keep terms to first order $\delta\xi^{\perp}_{x}$, we must keep terms to second order in $\Delta\chi_{x}$. Thus we have, 
\begin{equation}
T\left[ \chi ^{\prime }\right] =T\left[ \chi \right] +\int dx\frac{\delta T%
\left[ \chi \right] }{\delta \chi _{x}}\Delta \chi _{x}+\frac{1}{2}\int
dx\,dx^{\prime }\frac{\delta ^{2}T\left[ \chi \right] }{\delta \chi
_{x}\delta \chi _{x^{\prime }}}\Delta \chi _{x}\Delta \chi _{x^{\prime
}}+\cdots .  \label{Test Function b}
\end{equation}%
Substituting this expansion back into (\ref{Test Function a}) and computing the integral over $\chi^{\prime}$ using the transition probability (\ref{Trans Prob}), we obtain 
\begin{align}
\int D\chi \, \delta_{\chi} P[\chi|\chi_{0}]T\left[\chi \right]
=\int dx\,\frac{\eta \, \delta \xi^{\perp}_{x}}{g_{x}^{1/2}}&\int D\chi P[\chi|\chi_{0}]\notag\\
&\times\left\{ \frac{\delta T\left[ \chi \right] }{\delta \chi _{x}}\,\frac{\delta \phi \left[ \chi 
\right] }{\delta \chi _{x}}+\frac{1}{2}\frac{\delta ^{2}T\left[ \chi \right] }{\delta \chi _{x}^{2}}\right\}.
\label{Test Function c}
\end{align}%
To isolate and remove the arbitrary functional $T[\chi ]$ from both sides of (\ref{Test Function c}) we need to integrate by parts. Performing this move and discarding $T[\chi]$ we are left with
\begin{equation}
 \int dx \frac{\delta P[\chi|\chi_{0}]}{\delta\tilde{N}_{x}}\delta\xi_{x}^{\perp}=-\int dx\,\frac{\eta \, \delta \xi^{\perp}_{x}}{g_{x}^{1/2}}\frac{\delta }{\delta \chi _{x}}\left( P[\chi|\chi_{0}] \frac{\delta \phi \left[ \chi \right] }{\delta \chi _{x}}-\frac{1}{2}\frac{\delta}{\delta \chi _{x}}P[\chi|\chi_{0}]\right).
 \end{equation} 
Or, since $\delta \xi^{\perp}_{x}$ can be freely chosen, we obtain a finite separation local time Fokker-Planck equation
\begin{equation}
\frac{\delta P[\chi|\chi_{0}] }{\delta \tilde{N}
_{x}}=-\frac{\eta }{g_{x}^{1/2}}\frac{\delta }{\delta \chi _{x}}\left( P[\chi|\chi_{0}] \frac{\delta \phi \left[ \chi \right] }{\delta \chi _{x}}-\frac{1}{2}\frac{\delta}{\delta \chi _{x}}P[\chi|\chi_{0}]\right).
\end{equation}
If we apply the updating equation (\ref{Evolution equation}) for an arbitrary initial $\rho[\chi]=\rho_{\sigma}[\chi]$, then we obtain
\begin{equation}
\frac{\delta \rho _{\sigma }\left[ \chi \right] }{\delta \tilde{N}_{x}}=-%
\frac{1}{g_{x}^{1/2}}\left( \rho _{\sigma }\left[ \chi \right] \frac{\delta
\Phi _{\sigma }\left[ \chi \right] }{\delta \chi _{x}}\right) ~.
\label{FP Equation 2}
\end{equation}
with $\Phi_{\sigma}$ given by
\begin{equation}
\frac{\Phi _{\sigma }\left[ \chi \right]}{\eta} =\phi _{\sigma }\left[ \chi \right]- \log \rho _{\sigma }^{1/2}\left[ \chi \right],
\end{equation}%
which agrees with the LTFP equations (\ref{Local Time FP equation}).
\bibliographystyle{unsrtnat}

\end{document}